\documentclass[aps,reprint]{revtex4-1}

\usepackage{amsmath}
\usepackage{graphicx}
\usepackage{dcolumn}
\usepackage{verbatim}

\newcommand{\rhon}[1]{\rho^{(#1)}}
\newcommand{\phin}[2]{\phi_{#2}^{(#1)}}
\newcommand{\phins}[2]{\phi_{#2}^{*(#1)}}
\newcommand{\lphin}[2]{\langle\phi_{#2}^{(#1)}|}
\newcommand{\rphin}[2]{|\phi_{#2}^{(#1)}\rangle}
\newcommand{\nuks}{\nu_{ks}}
\newcommand{\nuksr}{\nu_{ks}(\mathbf{r})}
\newcommand{\nuksnn}[1]{\nu_{ks}^{(#1)}}
\newcommand{\nuksn}[1]{\nu_{ks}^{(#1)}(\mathbf{r})}

\newcommand{\ppnks}{\frac{\partial\nu_{ks}}{\partial\rho}}

\newcommand{\ppnksn}[1]{\frac{\partial^#1\nu_{ks}}{\partial\rho^#1}}
\newcommand{\br}{\mathbf{r}}
\newcommand{\bU}{\mathbf{U}}

\begin{document}

\title{Degenerate Density Perturbation Theory}
\author{Mark C. Palenik}
\thanks{NRC Research Associate}
\email{mark.palenik.ctr@nrl.navy.mil}
\author{Brett I. Dunlap}
\affiliation{Code 6189, Chemistry Division, Naval Research
Laboratory, Washington, DC 20375, United States}

\begin{abstract}
Fractional occupation numbers can be used in density functional theory to create a symmetric Kohn-Sham potential, resulting in orbitals with degenerate eigenvalues.  We develop the corresponding perturbation theory and apply it to a system of $N_d$ degenerate electrons in a harmonic oscillator potential.  The order-by-order expansions of both the fractional occupation numbers and unitary transformations within the degenerate subspace are determined by the requirement that a differentiable map exists connecting the initial and perturbed states.  Using the X$\alpha$ exchange-correlation (XC) functional, we find an analytic solution for the first-order density and first through third-order energies as a function of $\alpha$, with and without a self-interaction correction.  The fact that the XC Hessian is not positive definite plays an important role in the behavior of the occupation numbers.
\end{abstract}

\maketitle

\section{Introduction}
Quantum mechanical perturbation theory is used to build a map between eigenstates of two different Hamiltonians.  When the two Hamiltonians are similar, \textit{i.e.} they differ only by a small perturbing potential, it is possible to build a Taylor series in terms of the strength of the potential that connects the two wave functions and energies.

A problem arises when applying perturbation theory to a system that starts in a degenerate eigenstate.  The perturbation typically breaks the degeneracy and different linear combinations of initially degenerate states evolve into different perturbed states.  Any eigenstate arbitrarily chosen from the degenerate set is not guaranteed to evolve into eigenstate of the perturbed Hamiltonian.

In standard quantum mechanics, the solution to this problem is simple: diagonalize the perturbation within the degenerate subspace and choose one of the new basis vectors as the unperturbed state.  This is possible because the Hamiltonian is a linear operator, which means that any linear combination of two states with the same eigenvalue is also an eigenstate with the same eigenvalue.  At each order, perturbation theory can be solved by dividing the problem into two parts: one within the degenerate space and one in the orthogonal space.

In Kohn-Sham (KS) density functional theory (DFT) \cite{HKTheorems,Kohn1965}, on the other hand,  energy is not calculated from the eigenvalues of a linear operator, but as a functional of the electron density.  There is a similar eigenvalue problem, which can be divided into degenerate and nondegenerate subspaces, but the electrons are described in terms of single particle eigenstates of a non-linear operator.  This operator includes an effective potential, known as the KS potential, which is made up of electron-nuclear, Coulomb, and exchange-correlation (XC) parts, and is meant to reproduce the electron density of the interacting system.  The nonlinearity is contained within the Coulomb and XC potentials, which are used to model electron-electron interactions and are functions of the electron density.

It is the symmetry of the entire KS potential that determines the eigenvalue degeneracies.  Therefore, changing the unperturbed state can break the degeneracy even without an external perturbing potential.  If the system is open-shell, the density and correspondingly, the Coulomb and XC potentials, are symmetric if and only if each element of the open shell is occupied equally.  This can be accomplished with fractional occupation numbers, which are often necessary for the description of certain ground state densities, particularly when degeneracy is involved \cite{Levy1982,Schipper1998,Morrison2002,Katriel2004,Katriel1981,Englisch1984I,Englisch1984II}.  For example, a single electron that equally occupies three \textit{p} orbitals with occupation numbers of one third will produce a spherically symmetric density.



Because we have started with equal occupation numbers, we can diagonalize the perturbation to create a new set of basis states without altering the electron density.  However, each of the fractionally occupied states will have a different perturbed eigenvalue, leading to fractional occupation of excited orbitals.  If, on the other hand, we remove the fractional occupation numbers, the unperturbed electron density will be changed, destroying the symmetry of the KS potential.  Such a change requires a new self-consistent field (SCF) calculation, resulting in a new set of orbitals and eigenvalues. Perturbation theory, however, should be defined by a continuous connection between the unperturbed and perturbed states.

We derive the equations for degenerate perturbation theory by imposing the requirement of a continuous, differentiable map between the perturbed and unperturbed orbitals.  We will use an imaginary-time propagator to explicitly take this derivative.  The resulting equations necessitate an order-by-order change in occupation numbers, which maintains the eigenvalue degeneracy, provided the perturbation is small enough.  This is in line with a mathematical proof under mild assumptions which shows that two features of degenerate DFT perturbation theory are a change in natural occupation numbers at the Fermi level and a lack of eigenvalue splitting \cite{Cances2014}.  

At first glance, the lack of splitting would appear to be problematic for perturbation theory.  Typically differences between the formerly degenerate eigenvalues appear in the denominator of the equations that mix orbitals within the degenerate space at first order and higher \cite{sakurai2011modern}.  Level splitting induced by the perturbing potential would usually mean that these differences are no longer zero.  However, in DFT, because there is no splitting, these equations are singular.

However, the perturbing potential in DFT is not simply the external potential.  The KS potential changes at all orders, because it depends on the density, which also changes at all orders.  This order-by-order change in the Coulomb and XC potentials produces additional terms that prevent a singularity from occurring.

It may seem surprising that the perturbation does not lift the degeneracy, but this is only because at each order, we select a specific combination of orbital rotations and occupation numbers to make it so.  Despite the fact that the orbital eigenvalues remain degenerate, the perturbation singles out one particular state, which no longer has the unperturbed symmetry.

Physically, this can be understood by thinking of a two step process.  When the perturbing potential is applied, it raises the eigenvalues of some fractionally occupied orbitals higher than others.  Next, electrons transfer between orbitals until either the eigenvalues become degenerate again, or no more transfer of electrons is possible.

This rearrangement of electrons causes a change in the KS potential that restores the degeneracy of the initial state.  Such infinitesimal, degeneracy-preserving changes in the potential were investigated by Ullrich and Kohn, who found that that they are not rare and and are restricted by a relatively low number of conditions \cite{Ullrich2002}.

It is possible that adding a very small external potential could induce large changes in the Coulomb and XC potentials \cite{vanLeeuwen1994}, so that degeneracy cannot be restored by moving electrons between fractionally occupied orbitals in a physically viable manner.  However, the perturbed occupation numbers can still be meaningfully interpreted as derivatives with respect to the strength of the perturbing potential.  This holds so long as the XC functional is differentiable, thereby approaching its unperturbed value for an infinitesimal external perturbation.

If the fractionally occupied state represents an ensemble average of particles in the true, interacting system, then the eigenvalues should be independent of the occupation numbers.  Under this interpretation, the correct behavior of the energy is to linearly interpolate between the energy of different states as their occupation numbers are adjusted, causing the eigenvalues and XC potential to leap discontinuously at integer numbers of electrons \cite{Perdew1982,Sagvolden2008, Hodgson2016}.  It would then be impossible to equate two different eigenvalues by transferring electrons from one state to another.  However, this interpretation is not helpful when it comes to the SCF behavior of existing density functionals, and even seems to contradict the mathematical properties required for degenerate perturbation theory \cite{Cances2014}.

Existing density functionals are typically smooth, continuous functions of the electron density and do not display the discontinuities described by Perdew.  Such so-called $N$-continuous behavior is, additionally, a feature of the exact functional as defined by Cohen and Wasserman \cite{Cohen2003}, and in Landau's Fermi-liquid model \cite{landau1957, Thouless1972quantum, Nesbet1997}.  With these descriptions, the eigenvalues are free to vary in a continuous, differentiable manner as we transfer fractions of an electron from one orbital to another.


The change in occupation numbers can also be found by making the energy at a given order stationary with respect to the orbital occupations of a lower order.  This extends our work on density perturbation theory \cite{Palenik2015}, where we showed that the electron density at order $N$ can be found by varying the energy at order $N+M$ with respect to the density at order $M$.  Here we show that the same principle applies when occupation numbers are allowed to change.

From Janak's theorem \cite{Janak1978}, it can be seen that the fractional occupation numbers of a degenerate state extremize the energy when the number of Fermi-level electrons is held fixed.  The type of extremum (either a maximum, minimum, or saddle point) depends on matrix elements of the Coulomb plus XC Hessian within the degenerate space.  In general this Hessian is neither positive definite nor negative definite \cite{Dunlap2016}.  The occupation numbers at each order are proportional to its inverse, and the negative contribution of XC has a profound effect on their values.

We show this by analytically solving the perturbation equations for a system of electrons in a harmonic oscillator potential with the X$\alpha$ functional \cite{Slater1951exchange,Zope2005}. The lowest orbital contains two electrons of opposite spin and the first excited state is filled with up to three electrons of the same spin.  We find solutions both with and without a self-interaction correction (SIC) \cite{Perdew1981}, although such a correction is imperfect for fractional occupation numbers.  Adjusting the parameter $\alpha$ allows us to tune the amount of XC in the calculation.  Exchange correlation and the SIC both create a counterintuitive sign change that maximizes the interaction energy of the first-order density with the perturbing potential.  From the first-order density, we calculate the energy through third order, according to Wigner's $2n+1$ theorem \cite{Wigner1935,Angyan2009,Cances2014}.  By looking at this simple system, we demonstrate how perturbation theory can be applied when there is degeneracy.

\section{The degenerate perturbation problem in DFT}
Consider a mixed state described by a set of orbitals $\phi_i$ and occupation numbers $n_i$. We can write an expression for the energy in KS DFT as
\begin{equation}
\begin{split}
	E &= -\frac{1}{2}\sum_i n_i\langle\phi_i|\nabla^2|\phi_i\rangle \\
	&+ \int d\br\left[ V(\br)\rho(\br)+\int d\br'\frac{\rho(\br)\rho(\br')}{|\br-\br'|} +\varepsilon_{xc}[\rho(\br)]\right],
\end{split}
\label{KSEnergy}
\end{equation}
where $n_i$ is the occupation number of the $i$th orbital and $\varepsilon_{xc}[\rho]$ is the exchange-correlation energy-density as a function of the electron density.  The electron density, $\rho(\br)$, is equal to the sum of $n_i\phi^{*}_i(\br)\phi_i(\br)$ over all orbitals.

Making this energy stationary with respect to an orbital, $\phi_i^*$, given the constraint that the orbitals are orthonormal, yields the KS equation
\begin{equation}
	\frac{1}{n_i}\frac{\delta E}{\delta\phi_i^*}=\left(-\frac{1}{2}\nabla^2 + V(\br) + \nuks(\br)\right)|\phi_i\rangle = \epsilon_i|\phi_i\rangle,
	\label{EqEig}
\end{equation}
where $\nuks$ is the Coulomb plus exchange-correlation (XC) potential and $\epsilon_i$ is a Lagrange multiplier to enforce the constraint.  We will call the operator acting on $|\phi_i\rangle$ in Eq.~(\ref{EqEig}) $H_{KS}$, because it is like a quantum mechanical Hamiltonian.

Rayleigh-Schr\"odinger perturbation theory (RSPT) \cite{Shavitt2009} assumes the existence of a Taylor series connecting an initial solution of Eq.~(\ref{EqEig}) to a perturbed solution after the introduction of some potential $\lambda V^{(1)}(\br)$.  The orbitals, energy, and physical observables are then continuous, differentiable functions of the parameter $\lambda$ that scales the strength of the perturbing potential.  If such a Taylor series exists, the orbitals and KS operator, $H_{KS}$, can be expanded in powers of $\lambda$ as
\begin{eqnarray}
\phi_i = \phin{0}{i}+\lambda\phin{1}{i}+\lambda^2\phin{2}{i}+\ldots\\
H_{KS} = H_{KS}^{(0)}+\lambda V^{(1)} + \lambda \nuksnn{1}+\lambda^2\nuksnn{2}+\ldots
\end{eqnarray}
Unlike the external potential, $\nuks$ has terms at all orders of $\lambda$ because it depends on the density, $\rho$, which also changes at all orders.

By collecting the first-order terms from Eq.~(\ref{EqEig}) and taking the matrix element with another orbital, $\phin{0}{j}$, we can expand $\phin{1}{i}$ in terms of unperturbed orbitals.  Allowing $H_{ks}^{(0)}$ to act to the left creates a factor of $\epsilon_j^{(0)}$.  Rearranging terms, it is easy to show that if $i\neq j$, because of the orthonormality of the unperturbed orbitals,
\begin{equation}
\lphin{0}{j}V^{(1)}+\nuksnn{1}\rphin{0}{i} = (\epsilon_i^{(0)}-\epsilon_j^{(0)})\lphin{0}{j}\phi_{i}^{(1)}\rangle.
\label{EqEpsdiff}
\end{equation}

If orbitals $i$ and $j$ have degenerate eigenvalues, then the right hand side of Eq.~(\ref{EqEpsdiff}) is zero.  Under an arbitrary perturbation, the left hand side of Eq.~(\ref{EqEpsdiff}) is not zero, in general, and the usual solution in quantum mechanics is to create a new basis from linear combinations of degenerate states that diagonalize the first-order potential.  This diagonalizes the full Hamiltonian within the degenerate subspace and selects basis states that evolve into distinct eigenvalues when perturbed.

Of course, in DFT, the problem is slightly different because the first order potential includes $\nuksnn{1}$, which depends on the first order density and is given by
\begin{equation}
\nuksn{1}=\int\frac{\delta\nu_{ks}(\mathbf{r})}{\delta\rho(\mathbf{r'})}\rhon{1}(\mathbf{r'})d\mathbf{r'}.
\end{equation}
We will use the shorthand notation $\ppnks\rhon{1}$ to represent this integral.  Our expression for $\nuksn{1}$ comes from the first-order term in the Taylor series of $\nuks$ in $\rho$ around the unperturbed density.  Higher-orders terms can be found the same way \cite{Palenik2015}.

Because the degenerate orbitals are equally occupied, a unitary transformation between them does not change the density or KS potential.  Then, solving Eq.~(\ref{EqEpsdiff}) is only slightly more difficult than solving the equivalent equation in quantum mechanics.  The first-order density is found by solving the relevant coupled-perturbed Kohn-Sham equations \cite{Komornicki1993,Gonze1997,Fournier1990} or directly through density perturbation theory \cite{Palenik2015}.  Then, the total first-order KS potential can be diagonalized within the degenerate subspace.

However, the solution to Eq.~(\ref{EqEpsdiff}) is complicated by the fact that the first-order occupation numbers, which \textit{do} affect the density must change as well.  After the perturbing potential $V^{(1)}$ is applied and the degenerate levels split, several different perturbed excited states will all be equally occupied.  In that case, the energy is not stationary with respect to the fractional occupation numbers.

This problem can be solved by allowing the occupation numbers to change order by order.  Because changing the occupation numbers changes the KS potential, a small amount of level splitting can in some sense be ``canceled" by moving a fraction of an electron between previously degenerate states.  As the perturbation grows larger, assuming it breaks the symmetry, eventually, every state should either be fully occupied or empty and the occupation numbers will cease to change further.  However, it is still meaningful to talk about derivatives of the occupation numbers with respect to the perturbation parameter in the unperturbed state, and therefore, it is possible to build a complete perturbation theory.

In the following sections, we will define the requirements for a perturbation theory that takes the original fractionally occupied, degenerate state into a new fractionally occupied perturbed state.  These requirements define a coupled set of equations that determine the unitary transformations within the degenerate subspace and fractional occupation number changes at each order.

\section{Finding the lowest perturbed state}
\label{SecLowest}
In order for a Taylor series in $\lambda$ to exist connecting the unperturbed and perturbed states, at the very least, the unperturbed state must be differentiable with respect to $\lambda$.  To determine the requirements for differentiability, we will explicitly write an expression for the orbitals as a function of $\lambda$ and take its derivatives at $\lambda=0$.  We can then see where potential singularities may arise and how they can be corrected.
We will start by looking at a single orbital, but the advantage of this approach is that it will tell us how to treat fractionally occupied states as well.  
From here on, we will drop the zeroth-order superscript from the orbitals and eigenvalues, using $\phi_i$ and $\epsilon_i$ to refer to the $\lambda=0$ state.

We will derive time-independent perturbation theory starting from time-dependent perturbation theory.  Similar derivations are often used in quantum field theory to introduce interactions between fields.  Rather than propagating an orbital through real time, we can make a Wick rotation into imaginary time and propagate \textit{via} the equation $\phi_i(t)=e^{-H_{KS}t}\phi_i$.  Any orbital can be written as a linear combination of eigenstates, and under this time evolution the eigenstate with the lowest eigenvalue will decay the most slowly.   Therefore, normalizing this expression in the limit that $t$ goes to infinity leaves us with the lowest eigenstate that overlaps $\phi_i$.  Equivalently, this can be interpreted as taking the zero temperature limit of a thermal state.

When $H_{KS}$ is a function of $\lambda$, we then have an explicit expression for the lowest-eigenvalue orbital as a function of $\lambda$.  This will be true provided there is overlap with the original $\phi_i$ for all values of $\lambda$.  If there is not, a derivative discontinuity will occur.  For example, if $\phi_i$ starts in an excited state, the perturbing potential will very likely allow for mixing with a lower state, regardless of how weakly it is turned on.  Therefore, even under an infinitesimal but nonzero perturbation, $\phi_i$ will immediately transform into a lower state, and the derivative at $\lambda=0$ will be singular.

In practice, therefore, in our derivation, we will use the term ``lowest-eigenvalue orbital" to refer to an orbital occupied by an electron at the Fermi level.  Because the lower orbitals are fully occupied, we will only consider mixing with orbitals at the Fermi level or higher, effectively treating the lower lying states, for these purposes, as if they do not exist.

Before we consider fractional occupation numbers, let us look at an electron occupying a single orbital at the Fermi level.  We can propagate it through imaginary time and normalize it to find $\phi_i(\lambda)$, given by
\begin{equation}
|\phi_i(\lambda)\rangle=\lim_{t\rightarrow\infty}\frac{\sum_k|\phi_k\rangle\langle\phi_k|e^{-H_{KS}t}|\phi_i\rangle}{\sqrt{\sum_j|\langle\phi_j|e^{-H_{KS}t}|\phi_i\rangle|^2}},
\label{PhiG}
\end{equation}
where the indices $k$ and $j$ run over all orbitals at or above the Fermi level.  The choice of normalization in this case is not arbitrary.  Although in perturbation theory, the intermediate normalization is often convenient, where perturbed orbitals no longer have unit magnitude, it cannot be used DFT.  The energy functionals depend nonlinearly on the electron density, and it is therefore crucial that its magnitude, the number of electrons, is conserved.  This is ensured by the denominator of Eq.~(\ref{PhiG}).

Although the orbitals and therefore, the corresponding density, depend on $t$, the density that appears in $H_{KS}$ can be treated without approximation as a separate, time-independent quantity which is only a function of $\lambda$.  The operator $H_{KS}$ is then a function of the unperturbed density at $\lambda=0$ and the fully perturbed density at $\lambda=1$ for all values of $t$. All we require is that the time-dependent, orbital representation of the density becomes equal to the density in $H_{KS}$ in the limit that $t\rightarrow\infty$.  At earlier times, they need not be the same.

Separating the perturbed and unperturbed portions of the Hamiltonian, we can perform the time evolution of Eq.~(\ref{PhiG}) in the interaction picture.  To do so, we make the replacement $e^{-H_{ks}t}=\left(\mathcal{T}e^{-\int_0^t H'_{IP}dt}\right)e^{-H_{KS}^{(0)}t}$, where $\mathcal{T}$ is the time ordering operator and $H_{IP}'$ is the perturbation to $H_{KS}$ in the interaction picture, given by
\begin{equation}
H_{IP}' = e^{-H_{KS}^{(0)}t}\left(\lambda
V^{(1)}+\sum_{n=0}^{\infty}\lambda^n\nuksnn{n}\right)e^{H_{KS}^{(0)}t}.
\end{equation}

If the perturbation is weak, meaning that the perturbation does not cause the orbitals to change by much, the lowest
unperturbed eigenvector should evolve into the lowest perturbed eigenvector, so, we will choose this as $\phi_i$.  With this choice, $e^{-H_{KS}^{(0)}t}|\phi_i\rangle$ becomes $e^{-\epsilon_it}|\phi_i\rangle$ and this portion of the time evolution operator cancels in the numerator and denominator.  The orbital $\phi_i(\lambda)$ then can be written as
\begin{equation}
|\phi_i(\lambda)\rangle=\frac{\sum_k|\phi_k\rangle\langle\phi_k|\mathcal{T}e^{-\int_0^\infty
H'_{IP}dt}|\phi_i\rangle}{\sqrt{\sum_j|\langle\phi_j|\mathcal{T}e^{-\int_0^\infty
H'_{IP}dt}|\phi_i\rangle|^2}}.
\label{PhiGIP}
\end{equation}

By the definition of $H'_{IP}$, it can be seen that at $\lambda=0$, the time-evolution operator becomes the identity.  Then, the denominator becomes unity and the numerator becomes $\sum_k\delta_{ik}|\phi_k\rangle$.  Therefore, when the perturbation is turned off, our equation returns the initial, unperturbed state, as it must.

We can take derivatives of this state with respect to $\lambda$ at $\lambda=0$.  The first derivative is given by 
\begin{equation}
\begin{split}
\frac{d}{d\lambda}|\phi_i(\lambda)\rangle\big|_{\lambda=0} =
\sum_k|\phi_k\rangle\langle\phi_k|\frac{d}{d\lambda}\mathcal{T}e^{-\int_0^\infty
H'_{IP}dt}\big|_{\lambda=0}|\phi_i\rangle\\
-\sum_{j,k}\delta_{ik}\delta_{ij}|\phi_k\rangle\langle\phi_j|\frac{d}{d\lambda}\mathcal{T}e^{-\int_0^\infty
H'_{IP}dt}\big|_{\lambda=0}|\phi_i\rangle\\
=\sum_k(1-\delta_{ik})|\phi_k\rangle\langle\phi_k|\frac{d}{d\lambda}\mathcal{T}e^{-\int_0^\infty
H'_{IP}dt}\big|_{\lambda=0}|\phi_i\rangle.
\end{split}
\end{equation}

The first derivative of $\mathcal{T}e^{-\int_0^\infty H_{IP}'dt}$ at $\lambda=0$ is the term from its Dyson series proportional to $\lambda$, and we can therefore write an equation for the first-order perturbed orbital as
\begin{equation}
\rphin{1}{i}=-\sum_{k\neq i}|\phi_k\rangle\langle\phi_k|\int_0^\infty
e^{(\epsilon_i-\epsilon_k)t}\left(V^{(1)}+\nuksnn{1}\right)dt|\phi_i\rangle,
\label{Phi1Int}
\end{equation}
where $k$ is summed over all orbitals except for $i$.

It is clear that when there is no degeneracy, this integral produces the standard RSPT equations.  The $\epsilon_k$ eigenvalue is guaranteed to be greater than $\epsilon_i$ and therefore, the exponential vanishes at infinity and  Eq.~(\ref{Phi1Int}) becomes
\begin{equation}
\sum_{k\neq i}\frac{|\phi_k\rangle\langle\phi_k|V^{(1)}+\nuksnn{1}|\phi_i\rangle}{\epsilon_i-\epsilon_k}.
\end{equation}

When $\phi_k$ is degenerate with $\phi_i$, the exponential in Eq.~(\ref{Phi1Int}) becomes unity and the integral clearly no longer converges to a finite value.  If perturbation theory is to give a meaningful result, the matrix elements in Eq.~(\ref{Phi1Int}) must approach some well defined value as $t\rightarrow\infty$.  Another way of stating this is to say that the time-derivative of the matrix elements must go to zero as $t\rightarrow\infty$.  For the case of degenerate orbitals, setting the time-derivative of Eq.~(\ref{Phi1Int}) equal to zero gives us
\begin{equation}
\langle\phi_k|V^{(1)}+\nuksnn{1}|\phi_i\rangle=0,i\neq k,
\label{EqEpsDiffReprise}
\end{equation}
which is the same result that we found from the RSPT expansion in Eq.~(\ref{EqEpsdiff}).

Of course, we would like to extend this result to the more relevant case in DFT, when the degenerate orbitals are all fractionally occupied.  Because our system has started in the ground state, it makes sense that it should remain in the ground state after the perturbation is applied.  The fractional occupation numbers, if left unchanged, will cause each of the nondegenerate perturbed states to be equally occupied, and this must be remedied.

To do this, we will change the normalization of Eq.~(\ref{PhiGIP}).  The denominator, as we have currently written it, goes as $e^{-\epsilon_{\mathrm{min}}t}$ in the limit that $t\rightarrow\infty$, where $\epsilon_{\mathrm{min}}$ is the eigenvalue of the lowest lying perturbed state that overlaps $\phi_i$.  This has the effect of zeroing any terms in the numerator that decay faster than $e^{-\epsilon_{\mathrm{min}}t}$.

Let us simply call the eigenvalue belonging to the lowest lying perturbed state that overlaps any of the fractionally occupied orbitals $\epsilon$.  If we change the denominator so that it goes as $e^{-\epsilon t}$ in the limit that $t\rightarrow\infty$, this will have the effect of zeroing everything except for the ground state of the perturbed system.  To do so, we only have to add another summation to the denominator, so that Eq.~(\ref{PhiGIP}) becomes
\begin{equation}
|\phi_i(\lambda)\rangle=\frac{\sum_k|\phi_k\rangle\langle\phi_k|\mathcal{T}e^{-\int_0^\infty
H'_{IP}dt}|\phi_i\rangle}{\frac{1}{N_d}\sum_m\sqrt{\sum_j|\langle\phi_j|\mathcal{T}e^{-\int_0^\infty
H'_{IP}dt}|\phi_m\rangle|^2}},
\label{PhiGround}
\end{equation}
where $m$ is now summed over all degenerate orbitals and $N_d$ is the number of degenerate orbitals.  The factor of $1/N_d$ in the denominator comes from the fact that the initial occupation numbers are equal and conserves the total density within the degenerate space.  The derivative of $\phi_i(\lambda)$ with respect to $\lambda$ at $\lambda=0$ is, then
\begin{equation}
\begin{split}
\frac{d}{d\lambda}|\phi_i(\lambda)\rangle\big|_{\lambda=0} =
\sum_k|\phi_k\rangle\langle\phi_k|\frac{d}{d\lambda}\mathcal{T}e^{-\int_0^\infty
	H'_{IP}dt}\big|_{\lambda=0}|\phi_i\rangle\\
-\frac{1}{N_d}\sum_{j,k,m}\delta_{ik}\delta_{jm}|\phi_k\rangle\langle\phi_j|\frac{d}{d\lambda}\mathcal{T}e^{-\int_0^\infty
	H'_{IP}dt}\big|_{\lambda=0}|\phi_m\rangle\\
=-\sum_k|\phi_k\rangle\left[\langle\phi_k|\int_0^\infty
e^{(\epsilon_i-\epsilon_k)t}\left(V^{(1)}+\nuksnn{1}\right)dt|\phi_i\rangle\right.\\
 -\left. \frac{1}{N_d}\delta_{ik}\sum_j\langle\phi_j|\int_0^\infty\left(V^{(1)}+\nuksnn{1}\right)dt|\phi_j\rangle\right].
 \label{EqDegenElements}
\end{split}
\end{equation}
The index $j$ runs over degenerate orbitals.  Setting the time-derivative of each matrix element to zero in the limit that $t\rightarrow\infty$, for $\epsilon_i=\epsilon_k$, yields
\begin{equation}
\begin{split}
0=&\langle\phi_k|V^{(1)}+\nuksnn{1}|\phi_i\rangle\\
&-\frac{\delta_{ik}}{N_d}\sum_j\langle\phi_j|V^{(1)}+\nuksnn{1}|\phi_j\rangle.
\label{EqDegenElDeriv}
\end{split}
\end{equation}
For the diagonal elements, this implies that
\begin{equation}
\langle\phi_i|V^{(1)}+\nuksnn{1}|\phi_i\rangle=\frac{1}{N_d}\sum_j\langle\phi_j|V^{(1)}+\nuksnn{1}|\phi_j\rangle,
\end{equation}
or in other words, all of the first-order eigenvalues of the initially degenerate orbitals are equal.  Equation~(\ref{EqDegenElDeriv}) then can be written as
\begin{equation}
\langle\phi_k|V^{(1)}+\nuksnn{1}|\phi_i\rangle=\epsilon^{(1)}\delta_{ik},
\label{EqFirstOrder}
\end{equation}
where $\epsilon^{(1)}$ is the first order eigenvalue of all of the fractionally occupied orbitals.  We can continue to higher orders by taking additional derivatives of $\phi_i(\lambda)$ at $\lambda=0$.  By noting the fact that the $N$th term of a Taylor series has a factor of $1/N!$, we can see that the $N$th derivative of $\phi_i(\lambda)$ is $N!\phi_i^{(N)}$.  At each order, applying this procedure has the same effect of equating the eigenvalues.

\section{Solving for the perturbed density}
\label{SecSolve}
To solve for the perturbed density there are three things we will need to find: the transformation that mixes orbitals within the degenerate space, the mixing between occupied and virtual orbitals (for which no special degenerate techniques are required), and the first-order change in occupation numbers.  By solving Eq.~(\ref{EqFirstOrder}), we will show where these various quantities enter into the equations for perturbation theory and demonstrate how they can be found.  There are additional differences between DFT and standard quantum mechanics that become more apparent at second-order, and so, we will outline the second-order equations as well.

As is typically done in degenerate perturbation theory, it is useful to break up the problem into one part within the degenerate space and one within the orthogonal space.   A zeroth-order unitary transformation is required to diagonalize $V^{(1)}+\nuksnn{1}$ within the degenerate space, just as a zeroth-order unitary transformation is used to diagonalize $V^{(1)}$ in degenerate quantum mechanical perturbation theory.  This can be done without the need for a new SCF calculation because the degenerate orbitals are equally occupied, and so, this transformation leaves the unperturbed density unchanged.

Because this initial unitary transformation acts at zeroth order, we replace our initial basis of orbitals with a new basis given by
\begin{equation}
|\phi_i'\rangle = \sum_j \mathbf{C}^{(0)}_{ij}|\phi_j\rangle,
\end{equation}
with $i$ and $j$ summed over orbitals in the degenerate space.  Of course, at higher orders, the occupation numbers are not equal, so despite leaving the unperturbed density unchanged, this transformation will affect the perturbed density.

Let us factor $\mathbf{C}_{ij}^{(0)}$ out of the unitary transformation that mixes orbitals $i$ and $j$ at each order in perturbation theory and define the remainder as $\bU_{ij}$.  Expanding the full transformation in powers of $\lambda$, we get
\begin{equation}
\sum_k\bU_{ik}\mathbf{C}^{(0)}_{kj} = \sum_k\left(\delta_{ik}+\lambda\bU^{(1)}_{ik}+\lambda^2\bU^{(2)}_{ik}+\ldots\right)\mathbf{C}^{(0)}_{kj}.
\end{equation}
We can avoid explicitly writing $\mathbf{C}^{(0)}_{ij}$ in our equations by having $\bU_{ij}$ act on $\phi_j'$.

To see how $\bU_{ij}$ affects the density at each order, we can expand the density within the degenerate orbitals order-by-order to get
\begin{equation}
\begin{split}
&\sum_i\left[n_i\phi_i'^*\phi_i'\right]^{(N)}\\
&=\sum_{ijk}\sum_{l=0}^{N}\sum_{m=0}^{N-l}\sum_{p=0}^{N-l-m}\sum_{p=0}^{N-l-m-p}n_i^{(N-l-m-p-q)}\\
&\times\bU^{*(l)}_{ik}\bU^{(m)}_{ij}\phi_k'^{*(p)}\phi_j'^{(q)},
\end{split}
\end{equation}
where we are using $\phi_i'^{(q)}$ to represent the $q$th-order sum over states mixing of $\phi'_i$ with virtual orbitals

It can be seen that the term involving $\bU^{(N)}_{ij}$ is multiplied only by zeroth-order occupation numbers and orbitals.  Once again, because the zeroth-order occupation numbers are equal, a unitary transformation between them leaves the density unchanged.  Therefore, $\bU^{(N)}_{ij}$ does not affect the $N$th-order density.  It only comes into play at orders $N+1$ and higher.  Because of this, when we apply our equations to the Harmonic oscillator to find the first order density, which gives us the first through third-order energies, we will need to find $\mathbf{C}^{(0)}_{ij}$, but not $\bU^{(1)}_{ij}$.

Finding $\bU^{(N-1)}_{ij}$ is crucial for making an RSPT expansion possible at order $N$, and $\bU^{(N-1)}$ is a part of the $N$th-order perturbation problem, not the $N-1$th-order problem.  We will explicitly show that that solving for the first-order density determines $\mathbf{C}^{(0)}_{ij}$ and that the equations for the second-order second-order density determine $\bU^{(1)}_{ij}$.  This is not entirely different from standard quantum mechanics, where the equivalent first-order unitary transformation between degenerate states requires a second-order matrix element divided by first-order differences between eigenvalues.  However, in DFT, these eigenvalue differences are zero, and finding $\bU^{(1)}_{ij}$ will require simultaneously solving for all components of the second-order density.

Before moving on to second-order, however, we will show how the first-order density is found such that Eq.~(\ref{EqFirstOrder}) is satisfied.  The zeroth-order orbital rotation that diagonalizes the first-order potential is essentially the same as in standard quantum mechanics.  However, Eq.~(\ref{EqFirstOrder}) imposes the additional requirement that the first-order eigenvalues are equal.  For $N_d$ degenerate orbitals, equating the eigenvalues imposes $N_d-1$ new conditions on the solution to the perturbation equations. This requires $N_d-1$ additional adjustable parameters beyond the unitary transformation that diagonalizes $V^{(1)}+\nuksnn{1}$ in the degenerate subspace.  In total, the $1/2(N_d+2)(N_d-1)$ parameters needed to satisfy Eq.~(\ref{EqFirstOrder}) represent the restrictions on the first-order potential described by Ullrich and Kohn \cite{Ullrich2002} required to preserve degeneracy.

Fortunately, fractional occupation DFT has $N_d-1$ additional parameters that are not present in standard quantum mechanics, namely the first-order occupation numbers, which sum to zero, conserving the total number of electrons.  The occupation numbers at each order must independently sum to zero because they are defined, for all $\lambda$, in a series of the form $n_i^{(0)}+n_i^{(1)}\lambda+n_i^{(2)}\lambda^2+\ldots$.  In order for the number of electrons to be independent of $\lambda$, $\sum_in_i^{(0)}$ must provide the only nonzero contribution.

We can separate the portion of $\nuksnn{1}$ that is due the change in occupation numbers from the portion due to the change in orbitals and rewrite Eq.~(\ref{EqFirstOrder}) as
\begin{equation}
\begin{split}
\langle\phi_k'|V^{(1)}+\ppnks\left(\sum_j n_j^{(1)}\rho_j'^{(0)}+\rhon{1}_\phi\right)|\phi_i'\rangle=\epsilon^{(1)}\delta_{ik},
\end{split}
\label{EqN1Simple}
\end{equation}
where here, $j$ is summed over all occupied orbitals and $n_j^{(1)}$ is only nonzero at the Fermi level.  We are using the shorthand 
\begin{eqnarray}
&\rho_j'^{(0)}& = \phi'^*_j\phi_j'\\
&\rhon{1}_\phi&=2Re\sum_jn_j^{(0)}\phi_j'^{*(1)}\phi_j'=2Re\sum_jn_j^{(0)}\phins{1}{j}\phi_j.
\end{eqnarray}
Once again, because the $n^{(0)}_j$ are equal in the degenerate subspace, first-order mixing between these orbitals does not affect the first-order density.  For the same reason, $\mathbf{C}^{(0)}_{ij}$ does not affect $\rhon{1}_\phi$.  Terms with a prime depend on $\mathbf{C}^{(0)}_{ij}$, while unprimed terms do not.

The first-order eigenvalue, $\epsilon^{(1)}$, is $1/N_d$ times the trace of the left hand side of Eq.~(\ref{EqN1Simple}).  It is invariant under a unitary transformation on the orbitals because the terms proportional to $\rho_j'^{(0)}$ sum to zero.  This is because summing $n_j^{(1)}\langle\phi_i|\ppnks\phi_j^*\phi_j|\phi_i\rangle$ over all $i$ yields the expectation value of a symmetric operator, which must be invariant under a symmetry transformation.  The $n^{(1)}_j$ add up to zero to conserve the number of electrons.

Because the trace of the left-hand side of Eq.~(\ref{EqN1Simple}) is invariant under  a unitary transformation, the eigenvalue $\epsilon^{(1)}$, is then
\begin{equation}
\epsilon^{(1)}=\frac{1}{N_d}\sum_{i}\langle\phi_i|V^{(1)}+\ppnks\rhon{1}_\phi|\phi_i\rangle,
\end{equation}
which can be computed using the orbitals $\phi_i$, rather than $\phi_i'$.

Often, due to symmetry, $\langle\phi_k'|\ppnks\phi_j'^*\phi_j'|\phi_i'\rangle$ will be diagonal in $i$ and $k$, regardless of how the degenerate orbitals are rotated into each other.  For example, In the harmonic oscillator that we will explore later, the initial KS potential has spherical symmetry and the $\phi_i$ have odd parity.  Therefore, only the diagonal matrix elements are integrals of even functions and nonzero.  If the unitary transformation within the degenerate subspace decouples from the first-order occupation numbers in this manner, then Eq.~(\ref{EqN1Simple}) can be solved easily through an iterative process.

First, solve for the portion of $\rhon{1}$ that comes from the mixing of occupied and virtual orbitals with the assumption that $n^{(1)}_j=0$.  Although iteration is usually used at this step because the first-order KS potential is needed to find the first-order orbitals, it can be feasibly solved with a single matrix inversion for pure density functionals \cite{Palenik2015}.  Because the degenerate orbitals have equal zeroth-order occupation numbers, rotations within the degenerate subspace do not affect the density.  Therefore, we do not need to worry about the zeroth-order unitary transformation that diagonalizes $V^{(1)}+\nuksnn{1}$, nor do we need to worry about the first-order unitary transformation that mixes orbitals within the degenerate space.

Next, diagonalize $\langle\phi_i|V^{(1)}+\ppnks\rhon{1}|\phi_i\rangle$ to find a new set of orbitals $\phi_i'$.  We can then rewrite Eq.~\ref{EqN1Simple} as
\begin{equation}
\begin{split}
\sum_jn_j^{(1)}\langle\phi_i'|\ppnks\rho_j'^{(0)}|\phi_i'\rangle\\=\epsilon^{(1)}-\langle\phi'_i|V^{(1)}+\ppnks\rhon{1}_\phi|\phi_i'\rangle,
\end{split}
\label{EqN1LHS}
\end{equation}
Solving this equation for $n^{(1)}$ is a simple matter of inverting the matrix 
$\langle\phi_i'|\ppnks\rhon{0}_j|\phi_i\rangle$.  Explicitly inverting this matrix is not usually difficult because it has dimension equal to the number of degenerate orbitals, which is typically very small.  Now that a choice of $n^{(1)}$ has been made, we can go back to the first step and solve for $\rhon{1}_\phi$ again using the new first-order occupation numbers in the first-order KS potential.  This process can be iterated to self-consistency.

It should be noted that the diagonal matrix elements on the left-hand side of Eq.~(\ref{EqN1LHS}) represent self-interactions.  They come from the portion of the first-order KS potential due to the electrons occupying orbital $\phi_i$ acting on $\phi_i$.  The same is true of the terms of the right hand side that come from the first-order density due to the mixing of $\phi_i$ with virtual orbitals.  Therefore, a version of Eq.~(\ref{EqN1LHS})  with SIC \cite{Perdew1981} is
\begin{equation}
\begin{split}
\sum_jn_j^{(1)}\langle\phi_i'|\ppnks\rho_j'^{(0)}|\phi_i'\rangle(1-\delta_{ij})\\=\epsilon^{(1)}-\langle\phi'_i|V^{(1)}+\ppnks\left(\rhon{1}_\phi-2Re\phi_i'^{*(1)}\phi_i'\right)|\phi_i'\rangle.
\end{split}
\label{EqN1SIC}
\end{equation}

As Perdew and Zunger state, this SIC does not truly remove all self-interactions when fractional occupation numbers are used.  This becomes obvious when we consider the case of a single electron in the degenerate orbitals.  If self interactions are completely removed, the entire left hand side of Eq.~(\ref{EqN1SIC}) should be zero.  The left hand side represents interactions of the first-order density with the degenerate orbitals due to the first-order change in occupation numbers.  However, there is only a single electron within the degenerate space, and only this electron has first-order occupation numbers.

An alternate approach that we have taken to solving Eq.~(\ref{EqFirstOrder}), which may be more suitable when $\langle\phi_k|\ppnks\phi_j^*\phi_j|\phi_i\rangle$ is not guaranteed to be diagonal, is to employ a numerical optimization procedure to minimize
\begin{equation}
\sum_{ik}\left(\langle\phi'_k|V^{(1)}+\nuksnn{1}|\phi'_i\rangle-\epsilon^{(1)}\delta_{ik}\right)^2.
\label{EqMin}
\end{equation}
We do this by writing the matrix that mixes orbitals in the degenerate space as a function of $N_d(N_d-1)/2$ rotation angles.  We then minimize Eq.~(\ref{EqMin}) as a function of the total $(N_d+2)(N_d-1)/2$ rotation angles and first-order occupation numbers, setting $n^{(1)}_{N_d} = -\sum_{i=1}^{N_d-1}n_i^{(1)}$.

\subsection{The second-order density}
With the condition that the eigenvalue degeneracy is maintained, we find that the second-order RSPT expansion within the degenerate space is
\begin{equation}
\langle\phi_j'|V^{(1)}+\nuksnn{1}|\phi^{(1)}_i\rangle + \langle\phi_j'|\nuksnn{2}|\phi_i'\rangle=\epsilon^{(2)}\delta_{ij}+\epsilon_i^{(1)}\langle\phi_j'\rphin{1}{i},
\label{EqPhi2Degen}
\end{equation}
where $\epsilon^{(2)}$ is the second-order eigenvalue.  Although the first-order eigenvalues are equal as well, we will leave the index $i$ on $\epsilon_i^{(1)}$ for the time being.

Without the first-order mixing within the degenerate space, we would find that the term proportional to $\epsilon_i^{(1)}$ would be zero.   This means that all of the off-diagonal elements of the left-hand side of Eq.~(\ref{EqPhi2Degen}) would have to be zero as well.  In general, this is impossible.

If, on the other hand, we include the order-by-order unitary transformation that mixes orbitals within the degenerate space, we can break $\phi_i'^{(i)}$ into two parts, one which comes from $\bU^{(1)}_{ij}$ and one that comes from the usual sum over states that mixes occupied and virtual orbitals.  We then find that
\begin{equation}
|\phi_i'^{(1)}\rangle = \sum_{j}\bU^{(1)}_{ij}|\phi_j'\rangle +
\sum_a|\phi_a\rangle\frac{\langle\phi_a|V^{(1)}+\nuksnn{1}|\phi_i'\rangle}{\epsilon_i-\epsilon_a},
\end{equation}
where $j$ is summed over orbitals within the degenerate subspace and $a$ is summed over virtual orbitals.  Thus, the $\phi_a$ orbitals are unprimed.  Inserting this into Eq.~(\ref{EqPhi2Degen}), we get
\begin{equation}
\begin{split}
\sum_a\langle\phi_j'|V^{(1)}+\nuksnn{1}|\phi_a\rangle\frac{\langle\phi_a|V^{(1)}+\nuksnn{1}|\phi_i'\rangle}{\epsilon_i-\epsilon_a}\\
+\epsilon_j^{(1)}\bU^{(1)}_{ij}+\langle\phi_j'|\nuksnn{2}|\phi_i'\rangle=\epsilon^{(2)}\delta_{ij}+\epsilon_i^{(1)}\bU^{(1)}_{ij}.
\end{split}
\label{EqU1Init}
\end{equation}

The diagonal elements of $\bU^{(1)}_{ij}$, where $i=j$, are determined by the normalization we have chosen.  When there is no KS potential, the remainder of the elements can be found by grouping together the two terms that have an explicit factor of $\bU^{(1)}_{ij}$.  Solving these equations produces the usual formula from standard quantum mechanics,
\begin{equation}
\bU^{(1)}_{ij}=\sum_a\frac{\langle\phi_j|V^{(1)}|\phi_a\rangle}{\epsilon^{(1)}_i-\epsilon^{(1)}_j}\frac{\langle\phi_a|V^{(1)}|\phi_i'\rangle}{\epsilon_i-\epsilon_a}.
\end{equation}

This expression would be problematic for DFT perturbation theory, because while there is level splitting at first-order in quantum mechanics, in DFT, the eigenvalues remain degenerate at all orders, making the first denominator zero.  However, the KS potential introduces additional factors of $\bU^{(1)}_{ij}$, and Eq.~(\ref{EqU1Init}) can once again be solved.

The first-order unitary transformation within the degenerate space does not affect the first-order density.  Therefore, $\bU^{(1)}_{ij}$ does not factor into $\nuksnn{1}$.  However, it does affect the second-order density through the term $n^{(1)}_i\bU^{(1)}_{ij}$.  The full second-order density is given by
\begin{equation}
\begin{split}
\rhon{2} = \sum_j\left[n_j^{(2)}\rho'^{(0)}_j+2n_j^{(1)}Re\phi_j'^{*(1)}\phi_j'\right.\\
\left.+n_j^{(0)}\left(2Re\phi_j'^{*(2)}\phi_j'+\phi_j'^{*(1)}\phi_j'^{(1)}\right)\right].
\end{split}
\end{equation}
The unknown terms we must find to compute $\rhon{2}$ are $n_j^{(2)}$ and the factor of $\bU^{(1)}_{ij}$ contained within $2n_j^{(1)}Re\phi_j'^{*(1)}\phi_j'$.  Explicitly separating out their contribution to the second-order KS potential, we can rewrite Eq.~(\ref{EqPhi2Degen}) with all of the unknown quantities on the left and known quantities on the right as
\begin{equation}
\begin{split}
2Re\sum_{kl}\bU^{(1)}_{kl}n_k^{(1)}\langle\phi_j'|\ppnks\phi^*_k\phi_l|\phi_i'\rangle \\
+ \sum_jn_j^{(2)}\langle\phi_j'|\ppnks\rhon{0}_j|\phi_i'\rangle\\
=\epsilon^{(2)}\delta_{ij}-\sum_a\langle\phi_j'|V^{(1)}+\nuksnn{1}|\phi_a\rangle\frac{\langle\phi_a|V^{(1)}+\nuksnn{1}|\phi_i'\rangle}{\epsilon_i-\epsilon_a}\\
-\langle\phi_j'|\Delta\nuksnn{2}|\phi_i'\rangle,
\end{split}
\label{EqSecondOrderDensity}
\end{equation}
where $\Delta\nuksnn{2}$ is the remainder of $\nuksnn{2}$ that does not depend on $\bU^{(1)}_{ij}$ or $n^{(2)}$.  We can then solve Eq.~(\ref{EqSecondOrderDensity}) to find these unknown quantities, as we did with $\mathbf{C}^{(0)}_{ij}$ and $n_j^{(0)}$ at first-order.  A similar process can be performed at each order by finding the RSPT expansion of the orbitals, equating the eigenvalues, and solving for the $N$th-order occupation numbers and $N-1$th-order unitary transformations in the degenerate subspace.

The second-order occupation numbers are proportional to the same matrix that we needed to invert to find the first-order occupation numbers.  In fact, we will show that this is the case at all orders.  As we will see in the next section, this matrix is the Hessian of the electron-electron interaction energy given a fixed set of orbitals, or equivalently, the derivative of the $N$th-order eigenvalues with respect to the $N$th-order occupation numbers for $N>0$.

\section{Energy Extremization}
\label{SectionMin}
The fact that the levels do not split in the presence of a small, perturbing potential can be derived by making the occupation numbers at the Fermi level adjust themselves so as to extremize the energy.  According to Janak's theorem, the orbital eigenvalues are derivatives of the energy with respect to occupation numbers.  Therefore, moving electrons from an orbital with a higher eigenvalue into an orbital with a lower eigenvalue should lower the energy and vice versa.  However, unlike in quantum mechanics, where the Hamiltonian is linear, the nonlinear $H_{KS}$ operator of DFT depends on the orbital occupations.

Moving electrons from one orbital into another causes the eigenvalues to change.  Moving fractions of an electron from one orbital to another will only change the energy so long as the two eigenvalues remain different.  For orbitals that start out as degenerate, if a very weak perturbing field is applied with fixed occupation numbers, the eigenvalues should not split very much.  Therefore, it should be possible to make the eigenvalues equal again by moving a small fraction of an electron from one orbital to another.

The total number of electrons is always conserved, which places a constraint on the occupation numbers.  To extremize the energy at an arbitrary value of $\lambda$, we can write the equation
\begin{equation}
\frac{\partial E(\lambda)}{\partial n_j(\lambda)} = \epsilon(\lambda),
\end{equation}
where here, $\epsilon(\lambda)$ is a Lagrange multiplier to enforce the constraint.  By the application of Janak's theorem, the left-hand side is $\epsilon_j(\lambda)$.  Expanding both sides order-by-order in perturbation theory, we get $\epsilon_j^{(N)} = \epsilon^{(N)}$, meaning that the eigenvalues are equal at each order of perturbation theory as well.

More interestingly, we can make a connection to our prior work on density perturbation theory \cite{Palenik2015}.  There, we noted that in order to extremize the energy with respect to some variable (orbitals, density, \textit{etc.}) for all values of $\lambda$, the energy at each order must be extremized.  We also noted that the relevant variables at different orders are independent quantities.  Therefore energy at each order must be independently extremized with respect to the relevant variables at each order.  

When we chose the orbitals as the relevant variables, applying the calculus of variations to the energy order-by-order reproduced the standard RSPT perturbation series, in a manner similar to the method of Hylleraas \cite{Hylleraas1930}.  When we chose the density, expanded in a fitting basis, variation of the energy produced our density perturbation theory equations.  Here, we will apply this same principle to fractional occupation numbers.

Again, the total number of electrons is conserved, so we must extremize the energy with the constraint that the occupation numbers at first-order and higher sum to zero.  At the Fermi level, where fractional occupation numbers are allowed, we then get the equation
\begin{equation}
\frac{dE^{(N+M)}}{dn_j^{(M)}} = \epsilon^{(N)}.
\label{EqdEdn}
\end{equation}
Again, $\epsilon^{(N)}$ is a Lagrange multiplier.  It is clear that because the left hand side of Eq.~(\ref{EqdEdn}) has an $N+M$th-order term divided by an $M$th-order term, the equation must be of order $N$.  In fact, as one might expect from Janak's theorem, explicitly evaluating it produces the $N$th-order eigenvalue, $\epsilon_j^{(N)}$.  Therefore, we have once again equated all of the Fermi-level eigenvalues at each order.

When $N$ and $M$ are both zero, Eq.~(\ref{EqdEdn}) reduces to the standard form of Janak's theorem, but is problematic when $M$ is zero and $N$ is not.  The reason for this is that the zeroth-order orbitals and density depend self-consistently on one another.  Changing the zeroth-order occupation numbers changes the original SCF solution and therefore, is not really a part of perturbation theory.  However, for completeness, we will study the problem in more detail.

When $M$ is zero, Eq.~(\ref{EqdEdn}) includes terms equal to the derivative of the zeroth-order density with respect to the zeroth-order occupation numbers.  When $N$ is also zero, these terms either cancel or vanish due to the Hellman-Feynman theorem.  The self-consistent relationship between zeroth-order orbitals and occupation numbers leads to an implicit equation for $d\rho/dn_i^{(0)}$
\begin{equation}
\frac{d\rho(\br)}{dn_j^{(0)}} \mkern-2mu=\mkern-2mu \frac{\partial \rho(\br)}{\partial n_j^{(0)}}+2Re\sum_k\int \frac{\delta\rho(\br)}{\delta\phi_k(\br')}\frac{\delta\phi_k(\br')}{\delta\rho(\br'')}\frac{d\rho(\br'')}{dn_j^{(0)}}d\br'd\br'',
\end{equation}
or, grouping like terms,
\begin{equation}
\begin{split}
\int\left(\delta(\br-\br'')-2Re\sum_k \int\frac{\delta\rho(\br)}{\delta\phi_k(\br')}\frac{\delta\phi_k(\br')}{\delta\rho(\br'')}d\br'\right)\\
\times\frac{d\rho(\br'')}{dn_j^{(0)}}d\br''
= \frac{\partial \rho(\br)}{\partial n_j^{(0)}}.
\end{split}
\label{Eqdrhoimplicit}
\end{equation}
Applying Eq.~(\ref{EqdEdn}) when $N>0$ and $M=0$ requires solving this equation.

Variational fitting of the density to a series of auxiliary basis functions \cite{Dunlap2016} can be used to break this implicit dependence of $\rho$ on $\phi_i$.  Although $\phi_i$ is still a function of the density by way of the KS potential, this density is not a sum over orbitals squared, but an independent function, $\bar{\rho}$, that variationally minimizes the energy.  It is then possible to vary $E^{(N)}$ with respect to the zeroth-order $\bar{\rho}$ without solving Eq.~(\ref{Eqdrhoimplicit}) \cite{PalenikUnpublished}.

To avoid dealing with Eq.~(\ref{Eqdrhoimplicit}), we will demonstrate how we can generate an equation for $\epsilon^{(1)}$ by varying $E^{(2)}$ with respect to $n^{(1)}_j$.  Rather than using the form of the energy given in Eq.~(\ref{KSEnergy}), it is more convenient to use the well-known equivalent expression
\begin{equation}
E=\sum_i n_i\epsilon_i + \int \varepsilon_{ks}[\rho(\br)] - \nuksr\rho(\br) d\br,
\label{EqKSEnEig}
\end{equation}
where we have used $\varepsilon_{ks}[\rho]$ to represent the Coulomb energy-density, $\int \rho(\br)\rho(\br')/|\br-\br'|d\br'$, plus the XC energy density $\varepsilon_{xc}[\rho(\br)]$.   Expanding this to second order, we get
\begin{equation}
\begin{split}
E^{(2)}=\sum_i \left(n_i^{(0)}\epsilon_i^{(2)} + n_i^{(1)}\epsilon_i^{(1)}+n_i^{(2)}\epsilon_i\right)\\
 - \int \left[\nuks^{(2)}\rhon{0}(\br)+\frac{1}{2}\ppnks\rhon{1}(\br)\rhon{1}(\br)\right] d\br.
\end{split}
\label{EqKSEn2}
\end{equation}

From RSPT, the second-order eigenvalue, $\epsilon^{(2)}_i$ is \cite{Shavitt2009}
\begin{equation}
\langle\phi_i|V^{(1)}+\nuksnn{1}\rphin{1}{i} + \langle\phi_i|\nuksnn{2}|\phi_i\rangle,
\end{equation}
and the terms containing $\nuksnn{2}$ cancel from Eq.~(\ref{EqKSEn2}).  The derivative of $E^{(2)}$ with respect to $n^{(1)}_j$ is then
\begin{equation}
\begin{split}
\frac{\partial E^{(2)}}{dn^{(1)}_j} = \sum_i\left(n^{(1)}_i\langle\phi_i|\ppnks\phi_j^*\phi_j|\phi_i\rangle\right.\\
\left.+\langle\phi_i|\ppnks\phi_j^*\phi_j^*\rphin{1}{i} +\langle\phi_i|V^{(1)}+\nuksnn{1}\frac{d}{dn^{(1)}_j}\rphin{1}{i}\right)\\
+\epsilon^{(1)}_j-\int \ppnks\rhon{1}(\br)\phi_i(\br)\phi_i(\br).
\end{split}
\end{equation}
The RSPT expansion can be used to evaluate the derivative of $\phi_i^{(1)}$ with respect to $n_j^{(1)}$.  We then find that the third term on the right-hand side becomes
\begin{equation}
\begin{split}
\sum_a\langle\phi_i|V^{(1)}+\nuksnn{1}|\phi_a\rangle\frac{\langle\phi_a|\ppnks\phi_j\phi_j|\phi_i\rangle}{\epsilon_i-\epsilon_a} \\
= \lphin{1}{i}\ppnks\phi_j\phi_j|\phi_i\rangle.
\end{split}
\end{equation}

The first three terms on the right-hand side have factors of $\phi_i^*\phin{1}{i}$, $\phins{1}{i}\phi_i$, and $n_i^{(1)}\phi_i\phi_i$, which when added together and summed over $i$, gives us $\rhon{1}$.  These terms combine to make $\int\ppnks\rhon{1}(\br)\phi_i(\br)\phi_i(\br)$, which cancels the last term.

The only term remaining on the right hand side is $\epsilon_i^{(1)}$, which we set equal to the Lagrange multiplier $\epsilon^{(1)}$ to make the energy stationary.  This is exactly the expression we had previously arrived at equating the first-order eigenvalues.  The same equation can be produced, albeit with more difficulty, by varying $E^{(3)}$ with respect to $n^{(2)}_i$, $E^{(4)}$ with respect to $n^{(3)}_i$, and so forth.

We can note that if we were to differentiate with respect to the first-order occupation numbers a second time, we would be left with a zeroth-order quantity.  This quantity is the Hessian of the electron-electron interaction energy with respect to occupation numbers when the unperturbed orbitals are held fixed.  In other words, it is the second partial derivative of the energy with respect to occupation numbers, ignoring the dependence of the zeroth-order orbtitals on the zeroth-order occupation numbers. It, again, excludes the self-consistent relationship between the occupation numbers and density described by Eq.~(\ref{Eqdrhoimplicit}).  Because perturbation theory utilizes a basis of fixed, zeroth-order orbitals, this is the relevant Hessian for determining the behavior of the perturbed occupation numbers.  Differentiating the $N+M$th-order energy twice with respect to occupation numbers of order $M$ and $N$ will always produce this same matrix.  Equivalently, it can be obtained by differentiating the $N$th-order eigenvalues with respect to the $N$th-order occupation numbers.

This means that the $N$th-order eigenvalues are linear in the product of the $N$th-order occupation numbers and the Hessian of the electron-electron interaction energy.  Therefore, solving for the occupation numbers at any order requires the inverse of this matrix.

To find the Hessian, we can differentiate $E^{(2)}$ a second time with respect to $n^{(1)}_k$, and get
\begin{equation}
\frac{d^2E^{(2)}}{dn^{(1)}_jdn^{(1)}_k} = \frac{d\epsilon_j^{(2)}}{dn^{(1)}_k}=\langle\phi_j|\ppnks\phi^*_k\phi_k|\phi_j\rangle.
\label{EqEHess1}
\end{equation}
Dropping our shorthand notation $\ppnks$ for the moment, we can more easily see that these are matrix elements of the Coulomb plus XC energy Hessian, by rewriting Eq.~(\ref{EqEHess1}) as
\begin{equation}
\frac{d^2E^{(2)}}{dn^{(1)}_jdn^{(1)}_k} = \int \rhon{0}_j(\br)\frac{\delta\nuksr}{\delta\rho(\br')}\rhon{0}_k(\br')d\br d\br'.
\label{EqEHess2}
\end{equation}
The matrix $\delta\nuks(\br)/\delta\rho(\br')$ is the second derivative of the total electron-electron interaction energy with respect to the density.  The signs of its eigenvalues determine whether the energy is a maximum, minimum, or saddle point with respect to occupation numbers.  In the Hartree approximation, it is simply $1/|\br-\br'|$, which is positive definite, guaranteeing that the energy is minimized.  When XC is introduced, the Hessian has both positive definite and negative definite components, and this is no longer necessarily the case.  In Section~\ref{SecHarmonic}, this will be illustrated when we plot the behavior of the occupation numbers with respect to the parameter $\alpha$ in Slater's X$\alpha$ functional.

\section{The Harmonic Oscillator}
\label{SecHarmonic}
To test the properties of this perturbation theory, we apply it to a system of electrons in a harmonic oscillator potential.  Two electrons inhabit the lowest lying eigenstate with opposite spins and we vary the number of spin-up electrons in the first excited state, $N_d$, continuously between $0.05$ and $3$.  The unperturbed state is given spherical symmetry by equally occupying the degenerate states with occupation numbers of $N_d/3$.  Exchange and correlation are modeled by the X$\alpha$ functional, where the XC potential is proportional to $\alpha\rho^{1/3}$.  We perform our calculations both with and without the SIC described in Section~\ref{SecSolve}.

Because we are mostly interested in the behavior of the degenerate space, we will assume that the excitation energy is large compared to the strength of the perturbing potential so that we do not have to consider mixing with virtual orbitals.  This is acceptable for two reasons.  First, we are free to make the oscillator frequency, $\omega$ as high as we like, and in doing so, make level gap arbitrarily large.  Second, to cause mixing within the degenerate space, we need a perturbing field that couples orbitals of the same parity.  Therefore, any excited states that participate in mixing must be multiples of two levels higher.

The simplicity of this model allows us to find an analytic solution for occupation numbers as a function of $\alpha$.  Our solution can be taken to be exact in the limit that $\omega$ becomes infinite.  We treat the unperturbed ground state as independent of $\alpha$, which is also correct in this limit because the external potential provides the dominant interaction.


The solutions to the harmonic oscillator potential in three dimensions are products of Hermite polynomials in $x$, $y$, and $z$ times the Gaussian $e^{-\omega\br^2/2}$.  The zeroth Hermite polynomial, which represents the doubly-occupied lowest state, is $1$, and so, this orbital has spherical symmetry.  The first Hermite polynomials are simply $x$, $y$, and $z$.  Apart from the fact that there is only a single Gaussian exponent, these states are essentially the same as the basis functions used to represent \textit{s} and \textit{p} type orbitals in quantum chemistry.

Because the first Hermite polynomials are odd, for any spherically symmetric function $f(\mathbf{r})$, the matrix elements $\langle\phi_i|\phi_j\phi_kf(\mathbf{r})|\phi_l\rangle$ are only nonzero if there are two pairs of identical indices (\textit{e.g.} $i=j$ and $k=l$).  This is true even under a unitary transformation.  Therefore, matrix elements involving first-order occupation numbers, which have the form $n^{(1)}_j\langle\phi_k|\phi_j^*\phi_j|\phi_i\rangle$, are always diagonal.  This separates the first-order equation into two simpler problems: diagonalize $V^{(1)}$, then find the set of $n^{(1)}_j$'s that make the first-order eigenvalues equal.

As we showed in Section~\ref{SecSolve}, the first-order occupation numbers do not affect the first-order eigenvalue.  Because the occupied and virtual orbitals do not mix, the first-order eigenvalue is the trace of the perturbing potential in the degenerate space.  In the basis where $V^{(1)}$ is diagonal, we can then write an equation for $n^{(1)}_j$ that includes the SIC as
\begin{equation}
\begin{split}
n_j^{(1)}
=\left[\langle\phi_i'|\ppnks\rho_j'^{(0)}|\phi_i'\rangle(1-\delta_{ij})\right]^{-1}\\
\times\left[\frac{1}{N_d}\sum_k\langle\phi_k'|V^{(1)}|\phi_k'\rangle-\langle\phi_i'|V^{(1)}|\phi_i'\rangle\right],
\end{split}
\label{EqN1Solv}
\end{equation}
where the inverse should be taken to be a matrix inverse and the multiplication is matrix multiplication.  Without a SIC, the equation is the same, except that the $\delta_{ij}$ is removed.

A perturbing potential capable of mixing degenerate levels of the Harmonic oscillator must couple different states with the same parity. Neither a uniform electric field nor a dipole can do this, and so we chose a quadrupole field oriented along the $y$ and $z$ axes as our external perturbing potential, given by
\begin{equation}
V^{(1)}(\br) = \frac{Q(y^2-z^2)}{|\br|^5}.
\end{equation}

The unitary transformation required to diagonalize $V^{(1)}$ is simple, and trivial as we have set up the problem.  The quadrupole field has two coordinates that appear in it: $y$ and $z$, which represent any two arbitrary, orthogonal axes.  The Hermite polynomials are functions of three coordinates, $x$, $y$, and $z$, which could also represent any three arbitrary, mutually orthogonal axes.  The potential $V^{(1)}$ is diagonalized when the two axes that appear in the potential are oriented in the same direction as two of the axes used to define the Hermite polynomials.  Therefore, we shall define the basis of three degenerate orbitals $\phi_x = xe^{-\omega\br^2/2}$, $\phi_y = ye^{-\omega\br^2/2}$, and $\phi_z = ze^{-\omega\br^2/2}$.

The symmetries of $\phi_x$, $\phi_y$, and $\phi_z$ greatly reduce the number of independent matrix elements that need to be computed.  There is one independent, nonzero matrix element of $V^{(1)}$, two for exchange, and although there are three independent nonzero Coulomb integrals, only two actually appear in the Hessian.  The Coulomb and XC matrix elements are proportional to $\sqrt{\omega}$, while $V^{(1)}$ is proportional to $\omega^{3/2}$.  The exchange and $V^{(1)}$ matrix elements are additionally proportional to $\alpha$ and $Q$, respectively.

\begin{figure*}[t]
	\includegraphics[width=\columnwidth]{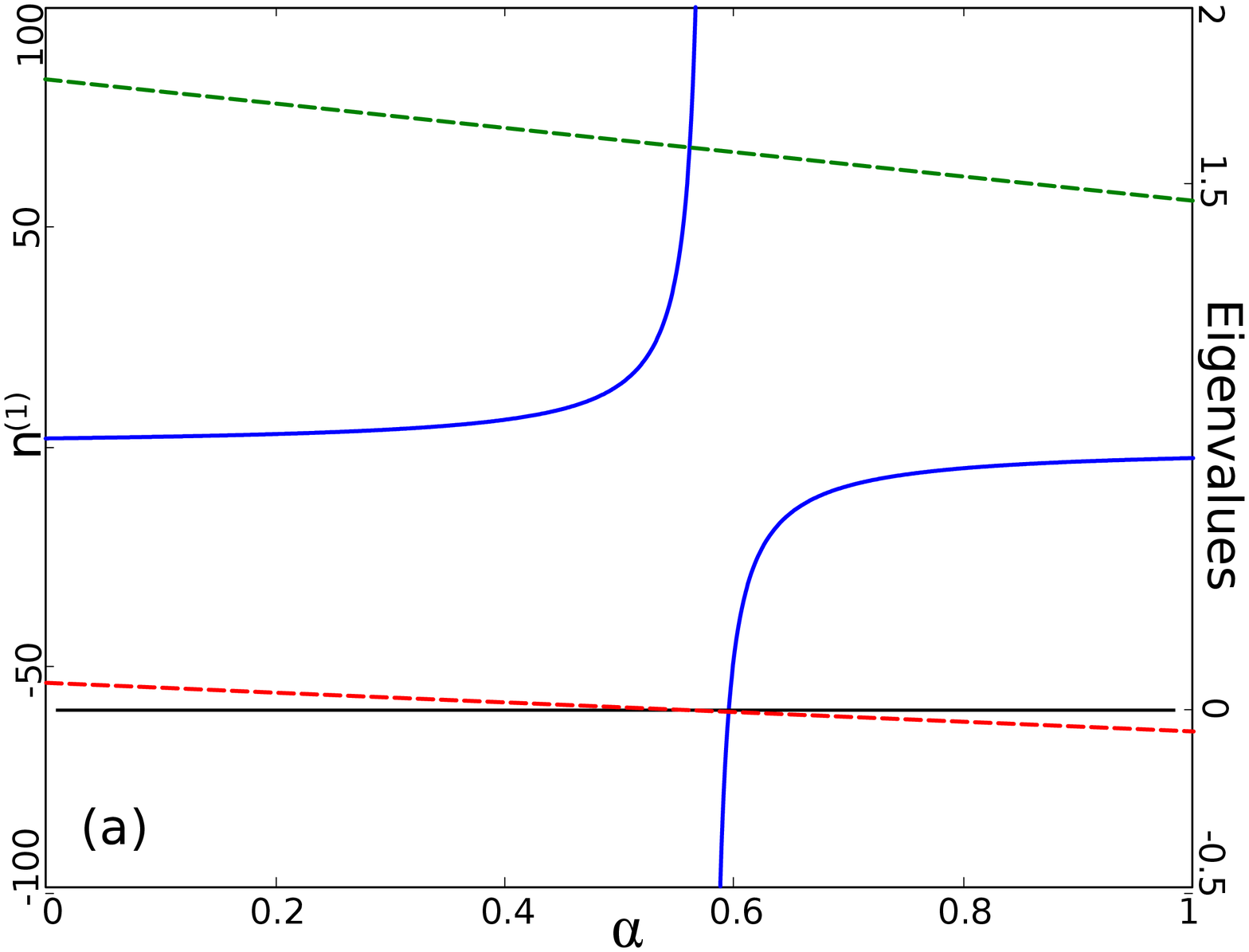}
	\includegraphics[width=\columnwidth]{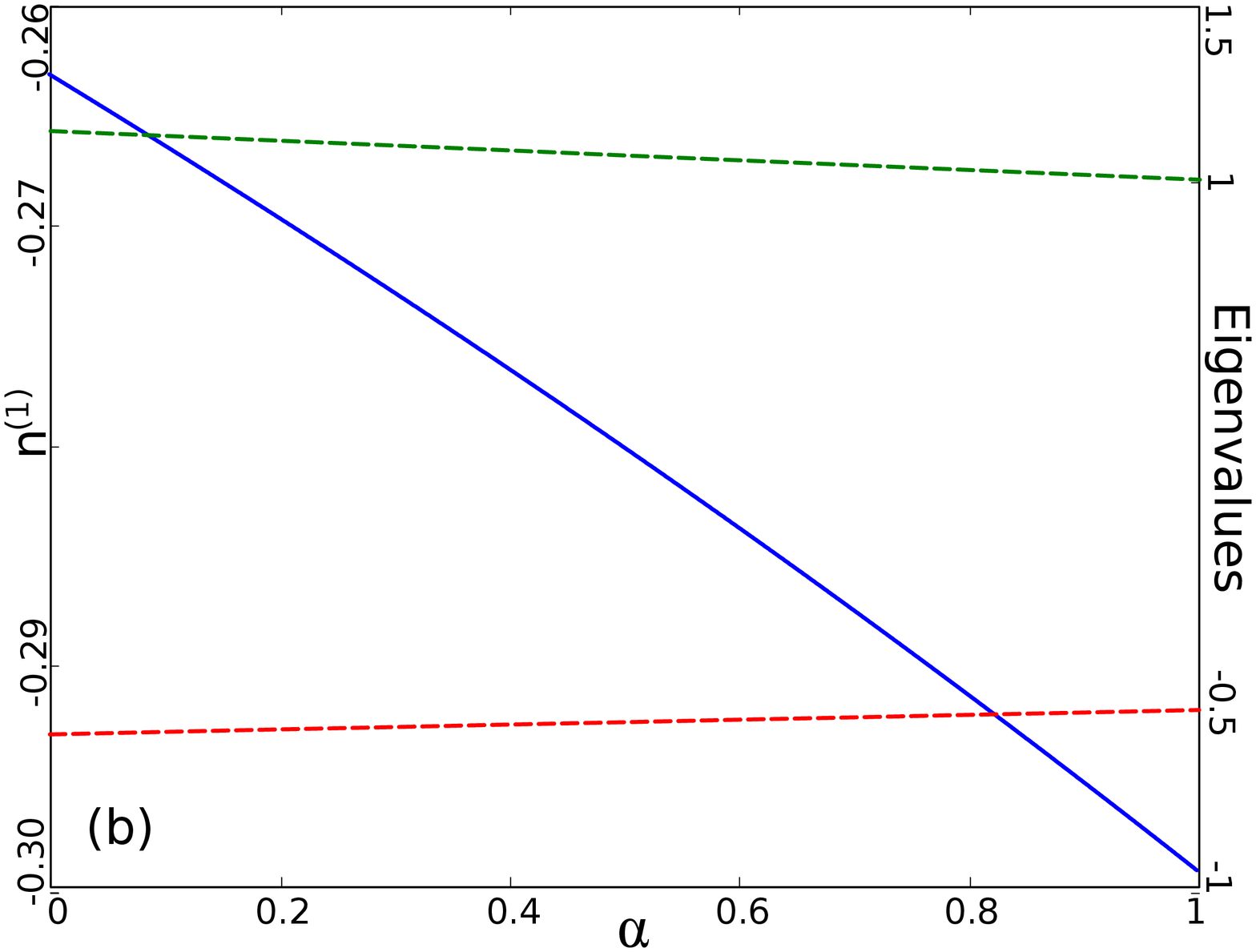}
	\caption{First order change in occupation numbers for one electron in the degenerate space when $\omega Q=1$ (solid lines) and Hessian eigenvalues (dashed lines) versus the parameter $\alpha$.  (a) without self-interaction correction (b) with self-interaction correction.  Note the vastly different scales on the vertical axis for $n^{(1)}$.  The horizontal black line in (a) highlights the eigenvalue zero-crossing.  The lower Hessian eigenvalue is degenerate.}
	\label{fignvsalpha}
\end{figure*}

\begin{figure}[h]
	\includegraphics[width=\columnwidth]{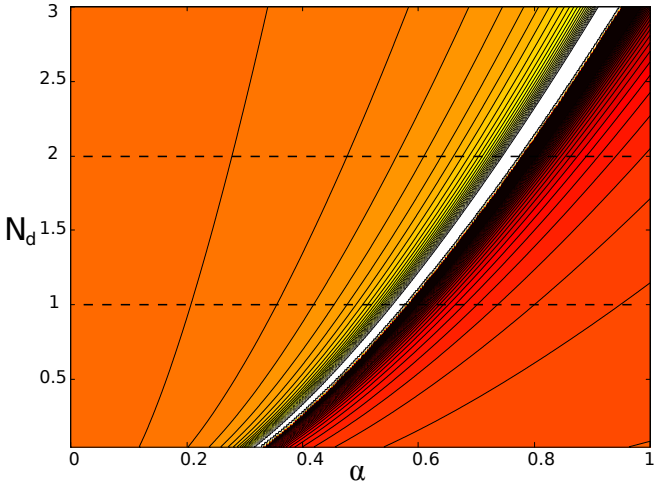}
	\caption{First-order fractional occupation numbers as a function of $\alpha$ and $N_d$.  $N_d$ ranges from $0.05$ to $3$.  Lighter colors represent higher values and darker colors are lower.}
	\label{fignvsNalpha}
\end{figure}

The trace of the quadrupole potential is zero, eliminating this term from Eq.~(\ref{EqN1Solv}).  The matrix elements of $V^{(1)}$, and therefore, first-order occupation numbers are linearly proportional to $\omega Q$.  The $\alpha$ dependence is more complicated, involving a quadratic function divided by a cubic function.

\begin{figure}[h]
	\includegraphics[width=\columnwidth]{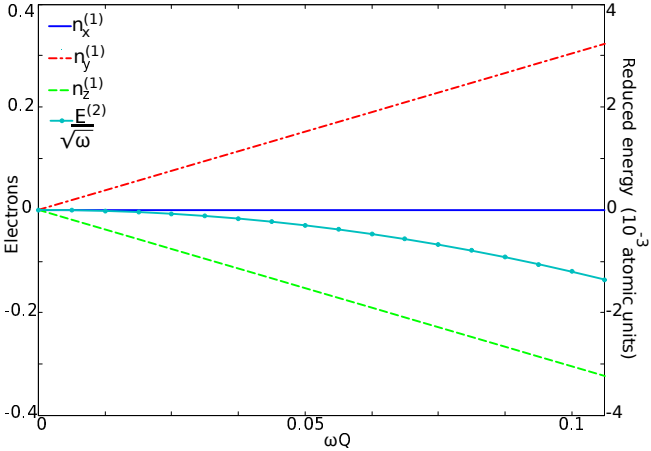}
	\caption{First-order fractional occupation numbers and second order energy versus $\omega Q$ at $\alpha=0.7$, without SIC.  The first and third-order energies are zero.  The second-order energy is proportional to $n^{(1)}_yn^{(1)}_z\sqrt{\omega}$.}
	\label{figurenvsq}
\end{figure}

The $\langle\phi_x|V^{(1)}|\phi_x\rangle$ and $\langle\phi_y|V^{(1)}|\phi_y\rangle$ matrix elements have opposite signs, while $\langle\phi_z|V^{(1)}|\phi_z\rangle$ is zero, and so, the same is true for $n^{(1)}_x$, $n^{(1)}_y$, and $n^{(1)}_z$ (Fig.~\ref{figurenvsq}).  Therefore, we can specify the first-order occupation numbers by a single parameter, $n^{(1)}$, which we will take to be $n^{(1)}_z$.

The Hessian, which is inverted to find the occupation numbers, becomes singular in two places when there is no SIC and in one place with a SIC.  This causes a singularity in the occupation numbers at all orders and is independent of the perturbing potential.  Without a SIC, the Hessian has two independent elements: the diagaonal terms and the off diagonal terms.  The XC contribution is linear in $\alpha$ for both, but with a different proportionality constant.  The first singularity occurs when the two values intersect, making every matrix element equal.  This corresponds to two of the three eigenvalues crossing zero, as can be seen in Fig.~\ref{fignvsalpha}(a), where we plot the first-order occupation numbers and Hessian eigenvalues versus $\alpha$ for one electron in the degenerate space.

With one electron in the degenerate space, this singularity occurs at $\alpha=0.577$ (a second also occurs at $\alpha=5.194$, when the third eigenvalue crosses zero).  In Fig.~\ref{fignvsNalpha}, we plotted $n^{(1)}$ versus $\alpha$ and $N_d$.  The Coulomb integrals are unaffected by $N_d$, and the only change to the equations is in the XC portion of the Hessian.  Because the XC energy goes as $\rho^{4/3}$, its second derivative goes as $\rho^{-2/3}$ and the XC integrals in the Hessian are approximately proportional to $(1+N_d)^{-2/3}$.

The singularity in Fig.~\ref{fignvsNalpha} is visible at the border between the black and white regions.  As $N_d$ is increased and the XC contribution becomes smaller, the value of $\alpha$ where the singularity occurs correspondingly increases.

\begin{figure*}[t]
	\includegraphics[width=0.655\columnwidth]{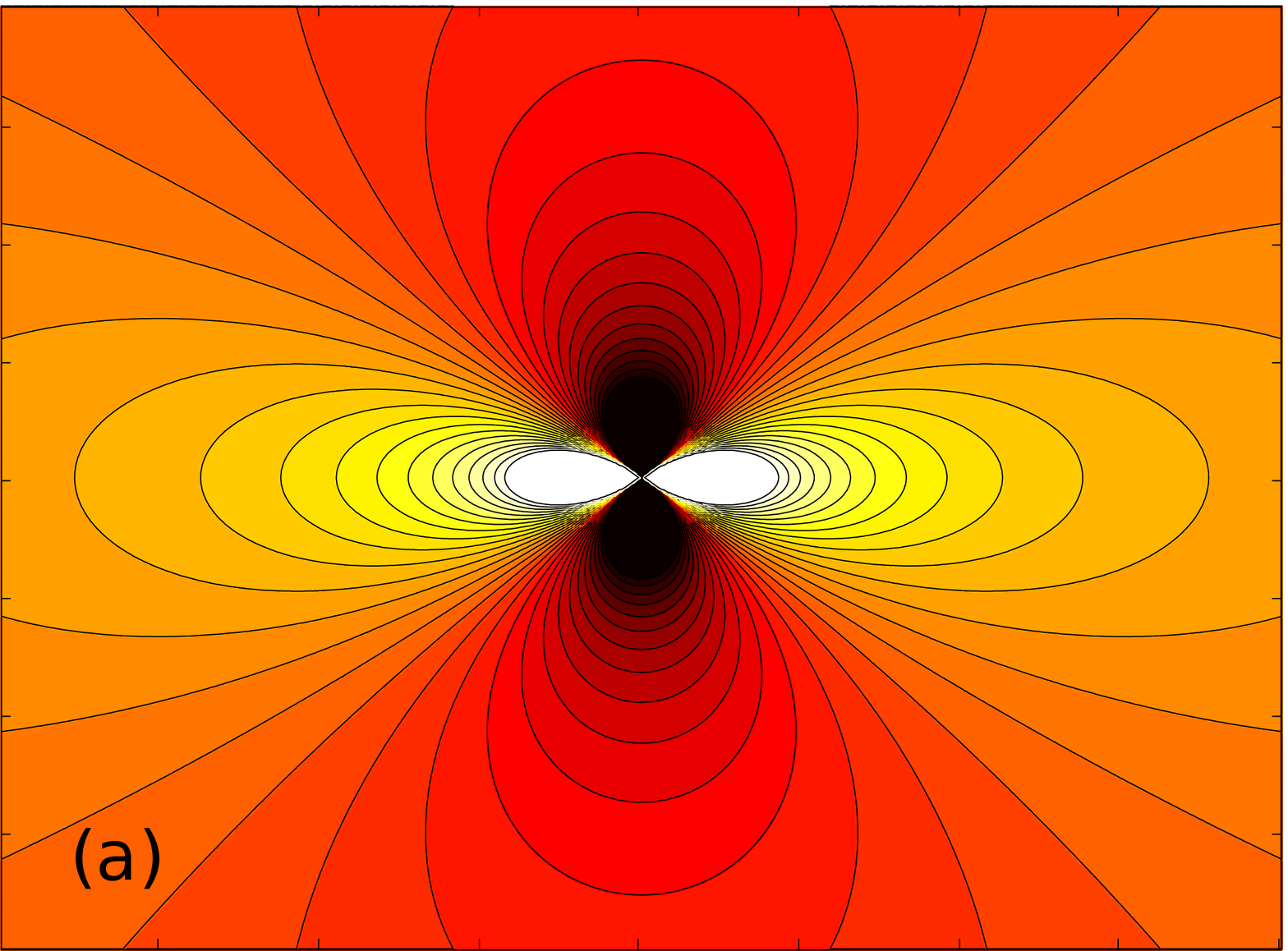}
	\includegraphics[width=0.66\columnwidth]{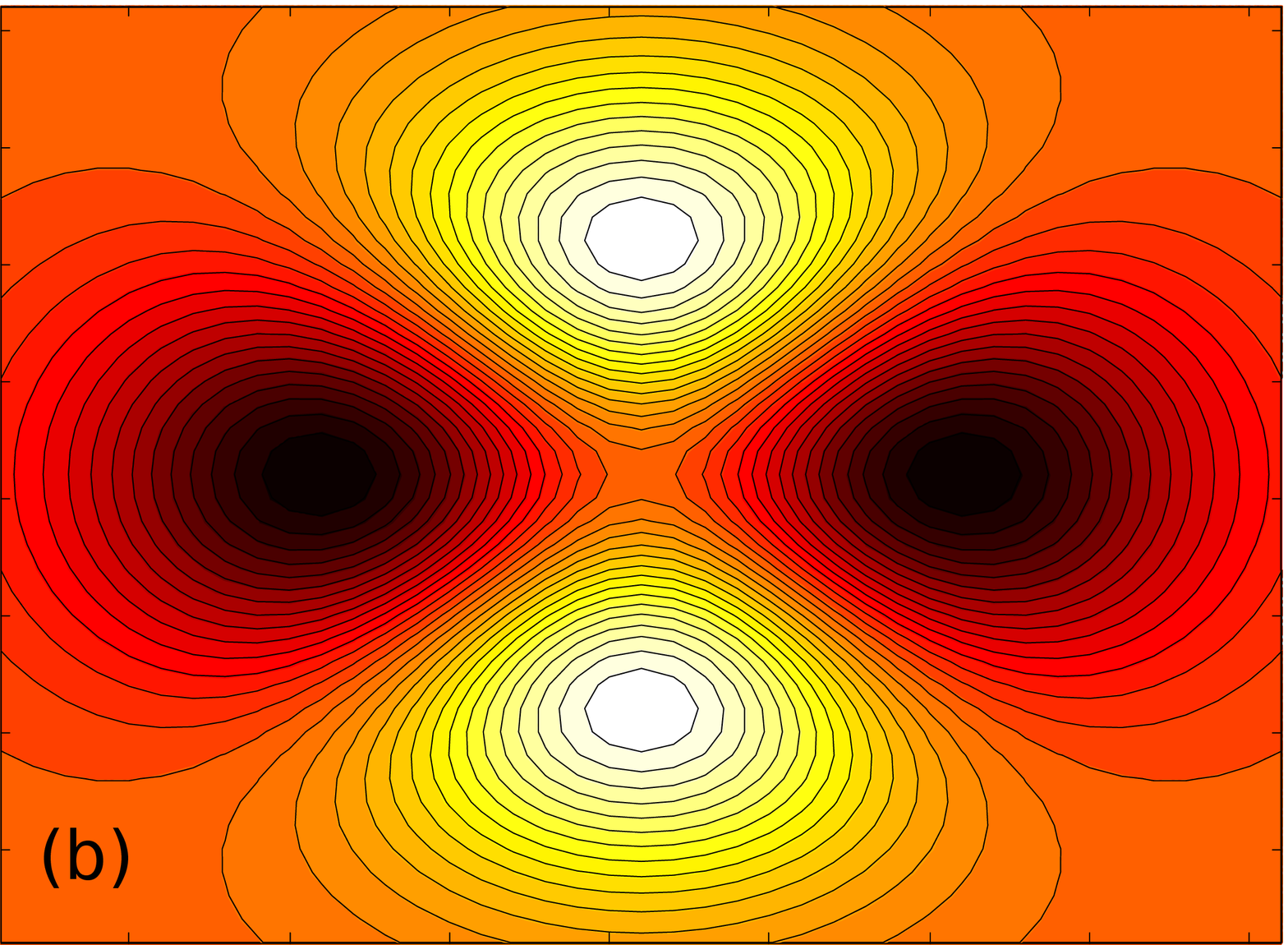}
	\includegraphics[width=0.66\columnwidth]{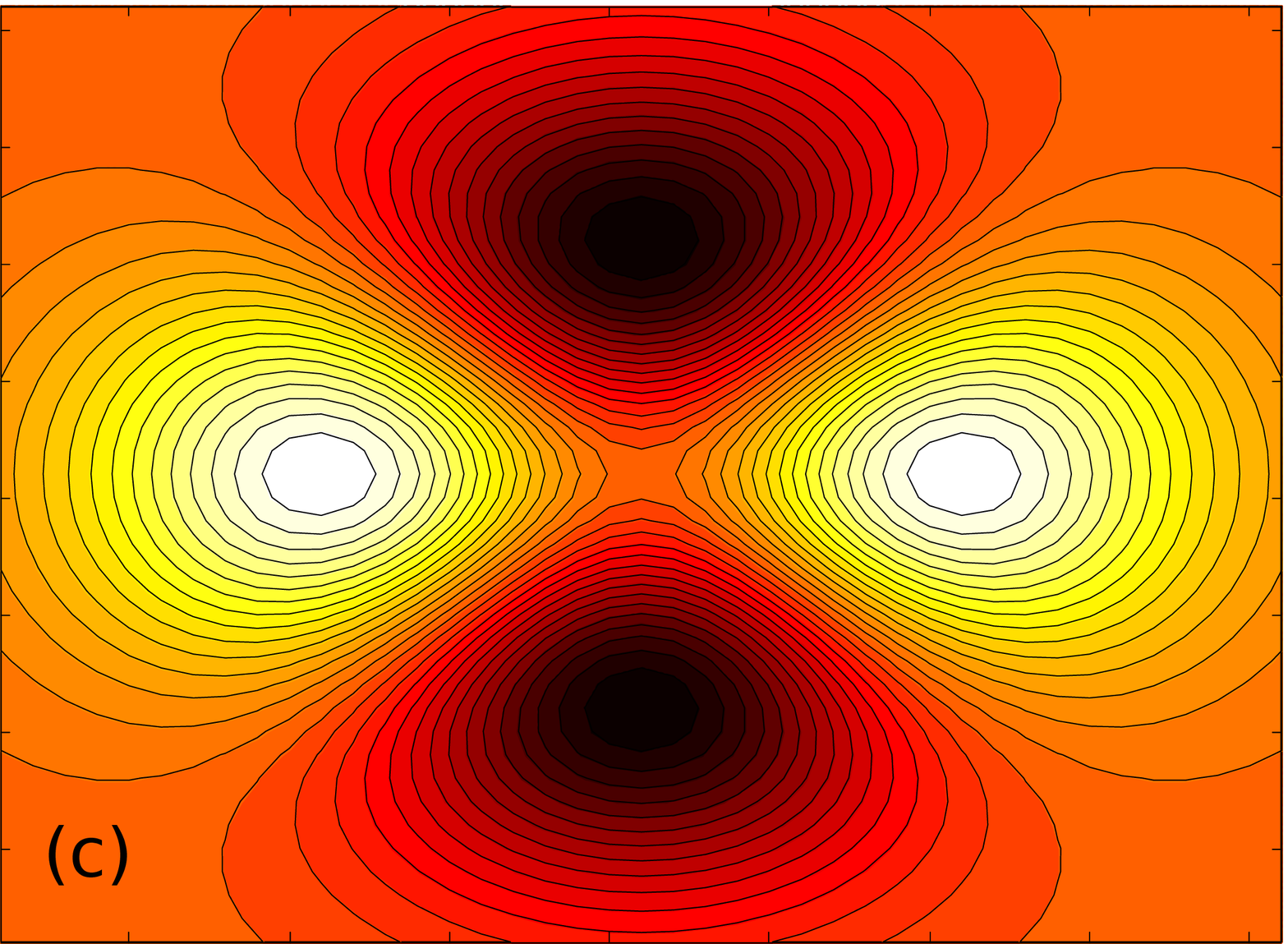}
	\caption{(a) perturbing external potential $V^{(1)}$ in the $yz$ direction (b) first-order density in the $yz$ direction when the Hessian is positive definite (c) first-order density in the $yz$ direction when the Hessian has negative eigenvalues.  Lighter colors represent higher values and darker colors are lower.}
	\label{figrhoandv}
\end{figure*}

Without a SIC, the Hessian initially has all positive eigenvalues within the degenerate subspace and the energy is minimized with respect to occupation numbers.  When two of the three eigenvalues simultaneously become negative, the first-order occupation numbers experience a counterintuitive sign change, as can be seen in Fig.~\ref{figrhoandv}.  There, we plotted the perturbing potential and first-order density along the $y$ (horizontal) and $z$ (vertical) axes.  In Fig.~\ref{figrhoandv}(c), where the Hessian has two negative eigenvalues, the first-order density is positive where the potential is positive and negative where the potential is negative, maximizing the interaction energy between the external perturbing potential and first-order density.  However, the total energy, which includes Coulomb and XC interactions, is not maximized.  The third Hessian eigenvalue is still positive, meaning the energy is extremized to a saddle point.

This saddle point extremization results in a negative $n^{(1)}$, meaning that electrons move from $\phi_z$ into $\phi_y$.  Because the $\phi_y$ matrix element of $V^{(1)}$ is positive and the $\phi_z$ matrix element is negative, we would expect moving electrons into $\phi_y$ to raise the energy.  This is not the case, however, because of the interdependence of each orbital eigenvalue on the occupation of all other orbitals.

Closer inspection shows that after the first eigenvalue zero-crossing, the off diagonal elements of the Hessian become greater than the diagonal elements.  This means that transferring electrons into any given orbital will raise the eigenvalue of other orbitals more than it raises its own.  Moving electrons from $\phi_z$ into $\phi_y$ actually corresponds to a Hessian eigenvector with a negative eigenvalue, meaning that the change in energy is negative to second order, as we will see when we explicitly calculate it.

Adding a SIC, which zeros the diagonal elements of the Hessian, gives the Hessian two negative eigenvalues over the entire range of $\alpha$ between zero and one.  It also has the effect of making $n^{(1)}$ much more uniform over this same range [Fig.~\ref{fignvsalpha}(b)].  The numerator and denominator of $n^{(1)}$ are both still quadratic and cubic functions of $\alpha$, respectively, but now they only have a single zero, at $\alpha=8.272$, when all of the matrix elements become zero


To first-order, the energy, as given by Eq.~(\ref{EqKSEnEig}) is
\begin{equation}
E^{(1)} = \sum_i\left(n_i^{(1)}\epsilon_i+n_i^{(0)}\epsilon_i^{(1)}\right) - \int\nuksn{1}\rhon{0}(\br)d\br.
\end{equation}
The first-order occupation numbers sum to zero and the eigenvalues are all equal, so the terms proportional to $n_i^{(1)}$ cancel.  The terms containing $\nuksnn{1}$ inside the first-order eigenvalues will also cancel with the integral, and so, the first order energy is
\begin{equation}
E^{(1)} = \sum_i\langle\phi_i|V^{(1)}|\phi_i\rangle=0.
\end{equation}

To find the second-order energy, we can start with the expression from Eq.~(\ref{EqKSEn2}).  Because we are not differentiating this expression, we can immediately cancel the terms proportional to $n_i^{(1)}$ and $n_i^{(2)}$, which sum to zero.  Because there is no mixing with virtual orbitals, the remaining terms can be simplified to
\begin{equation}
E^{(2)}=-\frac{1}{2}\int \ppnks\rhon{1}(\br)\rhon{1}(\br)d\br.
\end{equation}
The first-order density in our model problem is entirely determined by the first-order occupation numbers in the degenerate subspace.  Making the substitution $\rhon{1}=\sum_in_i^{(1)}\rhon{0}_i$, our final expression for the second-order energy, $E^{(2)}$, is
\begin{equation}
E^{(2)} = -\frac{1}{2}\sum_{ij} n_i^{(1)}n_j^{(1)}\int \ppnks\rhon{0}_i(\br)\rhon{0}_j(\br)d\br.
\label{EqE2Final}
\end{equation}
For our problem, this energy is quadratic and decreasing in $n^{(1)}$.  The integral in Eq.~(\ref{EqE2Final}) is proportional to $\sqrt{\omega}$, and so in Fig.~\ref{figurenvsq}, we plot $E^{(2)}$ divided by $\sqrt{\omega}$ versus $\omega Q$.

By the $2n+1$ theorem \cite{Wigner1935,Angyan2009,Cances2014}, we can also find the third-order energy from the first-order density.  In this case, it will turn out to be zero.  Expanding the energy as defined in Eq.~(\ref{EqKSEnEig}) to third-order and simplifying the terms inside the integral that come from $\varepsilon_{xc}[\rho(\br)]$ and $\nuksr\rho(\br)$, we get
\begin{equation}
\begin{split}
E^{(3)}=\sum_i\left[n_i^{(0)}\epsilon_i^{(3)}+n_i^{(1)}\epsilon_i^{(2)}+n_i^{(1)}\epsilon_i^{(2)}+n_i^{(3)}\epsilon_i\right]\\
-\int\left[\nuksn{3}\rhon{0}(\br)+\ppnks\rhon{2}(\br)\rhon{1}(\br)\right.\\
\left.+\frac{1}{3}\ppnksn{2}\rhon{1}(\br)\rhon{1}(\br)\rhon{1}(\br)\right]d\br.
\end{split}
\label{EqE3}
\end{equation}
Again, the terms proportional to $n_i^{(1)}$, $n_i^{(2)}$, and $n_i^{(3)}$ will sum to zero.  The third-order eigenvalue, $\epsilon^{(3)}$, which is the same for all degenerate orbitals, is given by
\begin{equation}
\begin{split}
\epsilon^{(3)} = \langle\phi_i|V^{(1)}+\nuksnn{1}\rphin{2}{i} + \langle\phi_i|\nuksnn{2}\rphin{1}{i} \\
+\langle\phi_i|\nuksnn{3}|\phi_i\rangle+ \frac{\epsilon^{(1)}}{2}\lphin{1}{i}\phi_i^{(1)}\rangle.
\end{split}
\label{EqEps3}
\end{equation}
The last term on the right comes from the normalization we have chosen, which sets $\langle\phi_i\rphin{2}{i}=-1/2\lphin{1}{i}\phi_i^{(1)}\rangle$, but this term disappears, regardless, because $\epsilon^{(1)}=0$.

Because we are working in the limit that $\omega$ becomes infinite, $\phi_i^{(1)}$ and $\phi_i^{(2)}$ only include the unitary transformation within the degenerate space.  This causes the first term on the right of Eq.~(\ref{EqEps3}) to be zero, because it becomes $\sum_{j}\bU^{(2)}_{ij}\langle\phi_i|V^{(1)}+\nuksnn{1}|\phi_j\rangle=\epsilon^{(1)}\delta_{ij}=0$.

The final remaining term in $\epsilon^{(3)}$ must also contribute nothing to $E^{(3)}$.  The second order potential $\nuksnn{2}$ is sandwiched by the orbitals $\phi_i^*\phin{1}{0}$, which can be written as $\sum_{i}\bU^{(1)}_{ij}\phi^*_i\phi_j$.  The sum $\sum_{ij}\bU^{(1)}_{ij}\phi^*_i\phi_j$ must be equal to zero, because $\bU_{ij}$ is a unitary transformation, which leaves $\sum_i\phi_i^*\phi_i$ unchanged at all orders and $\epsilon$ is real.

Finally, we can note that the $\nuksnn{3}$ term in $\epsilon^{(3)}$, when summed over all $i$ will cancel with the $\nuksnn{3}$ term in Eq.~(\ref{EqE3}). We can then write the third-order energy as
\begin{equation}
\begin{split}
E^{(3)} = -\int \left(\ppnks\rhon{2}(\br)\rhon{1}(\br)\right.\\
\left.+\frac{1}{3}\ppnksn{2}\rhon{1}(\br)\rhon{1}(\br)\rhon{1}(\br)\right)d\br.
\end{split}
\end{equation}

All that is left is to eliminate the second-order density from $E^{(3)}$.  This can be done from the definition of $\epsilon^{(2)}$.  Rearranging terms, we get
\begin{equation}
\begin{split}
\int \ppnks\rhon{2}(\br)\rhon{0}_i(\br)d\br \\
= \epsilon^{(2)} - \frac{1}{2}\int\ppnksn{2}\rhon{1}(\br)\rhon{1}(\br)\rhon{0}_i(\br)d\br.
\end{split}
\label{EqRhon2Def}
\end{equation}
Noting that $\rhon{1}=\sum_in_i^{(1)}\rhon{0}_i$, we can multiply both sides of Eq.~(\ref{EqRhon2Def}) by $n_i^{(1)}$ and sum over $i$.  Because the $n_i$ sum to zero, the term proportional to $\epsilon^{(2)}$, which is independent of $i$, will cancel out.  Our final expression for $E^{(3)}$, is then
\begin{equation}
\begin{split}
E^{(3)} = \frac{1}{6}\int\ppnksn{2}\rhon{1}(\br)\rhon{1}(\br)\rhon{1}(\br)d\br\\
=\frac{1}{6}\sum_{ijk}n^{(1)}_in^{(1)}_jn^{(1)}_k\int\ppnksn{2}\rhon{0}_i(\br)\rhon{0}_j(\br)\rhon{0}_k(\br)d\br.
\end{split}
\label{EqE3Final}
\end{equation}
The presence of symmetries, namely that $n^{(1)}_y=-n^{(1)}_z$ and $n^{(1)}_x=0$, along with the fact that permutations of integrals involving $\rhon{0}_y\rhon{0}_y\rhon{0}_z$ and $\rhon{0}_z\rhon{0}_z\rhon{0}_y$ are equal, means that the sum of all terms in Eq.~(\ref{EqE3Final}) is zero.  Therefore, our result simply reduces to $E^{(3)}=0$.

\section{Conclusions}
Fractional occupation numbers can be used to give the total KS potential the same symmetry as the nuclear potential.  This results in a set of fractionally occupied, degenerate orbitals at the Fermi level. We have derived the equations for the perturbation theory that results when a symmetry breaking external potential is applied that couples the degenerate states and applied them to a simple model of electrons in a harmonic oscillator potential.  By working in the limit that $\omega$, the oscillator frequency, becomes infinite, we are able to arrive at an analytic solution for the first-order density and first-through third order energy as a function of $\omega$ and the parameter $\alpha$ in the X$\alpha$ functional.

The inclusion of exchange has a profound effect on first-order occupation numbers. In the Hartree approximation, when there is no exchange and no SIC, the Hessian of the electron-electron interaction energy is positive definite.  This means that the energy is minimized with respect to occupation numbers.  Exchange and correlation add a non-positive definite portion to the Hessian, and a minimization is no longer guaranteed.  Without the SIC, the Hessian within the subspace of degenerate orbitals is positive definite at $\alpha=0$.  With one electron in the degenerate orbitals, the Hessian becomes singular at $\alpha=0.577$, where the first-order occupation numbers also become singular and then change sign.

The number of electrons in the degenerate space, $N_d$, only affects the equations for perturbation theory through the XC contribution to the Hessian.  The greater $N_d$, the smaller the magnitude of the XC integrals, because the second derivative of the XC energy goes as $\rho^{-2/3}$.  Therefore, increasing $N_d$ pushes the singularity to larger values of $\alpha$.

Including the SIC zeros the diagonal elements of the Hessian.  This gives it negative eigenvalues even at $\alpha=0$.  There is no sign change in the first-order occupation numbers as $\alpha$ goes from zero to one, and only approximately a thirteen percent change in their values over this range.

We found that the Fermi-level occupation numbers must change to preserve the degeneracy of the perturbed system, in agreement with theorems proven by Canc\`es and Mourad \cite{Cances2014}.  This is related both to the Janak theorem \cite{Janak1978} and our previous work in density perturbation theory \cite{Palenik2015}, where we showed that the energy at each order is made stationary by the density at all other orders.

Because DFT involves eigenstates of a nonlinear operator, the orbital eigenvalues depend on the occupation numbers.  Shifting electrons between two states with different eigenvalues causes both eigenvalues to change.  In the limit of a small perturbing potential that breaks the unperturbed degeneracy, such changes can continue until the eigenvalues become equal once again.  If the potential is large enough, eventually, either some states will become completely empty, or others will be fully occupied, and the degeneracy will begin to lift.

In all of our calculations, when $\alpha$ was between zero and one, every matrix element of the Hessian was positive.  The positive elements along the diagonal mean that transferring an electron into a given orbital raises the eigenvalue of that same orbital.  However, the eigenvalues also depend on the occupation of all other orbitals in the degenerate space, as determined by the off-diagonal elements of the Hessian.  When the off-diagonal elements are larger than the diagonal elements, the Hessian is no longer positive definite and there is some combination of electron transfers that lowers the energy.  In this scenario, both the perturbed and unperturbed states are energy extrema, but not minima.

Our equations for perturbation theory stem entirely from the requirements that a differentiable map exists which connects eigenstates of the perturbed and unperturbed systems.  We explicitly build this differentiable map by using an imaginary time propagator to find the ground state as a function of the parameter $\lambda$, which scales the strength of the perturbing field.  By differentiating this expression, we are able to find the order-by-order equations for both occupation numbers and orbital rotations within the degenerate subspace that are needed to make perturbation theory work.

\begin{acknowledgments}
This work is supported by the Office of Naval Research, directly and through the Naval Research Laboratory. M.C.P. gratefully acknowledges an NRC/NRL Postdoctoral Research Associateship.
\end{acknowledgments}

\bibliography{citations}{}
\end{document}